\documentclass[twocolumn,journal]{IEEEtran}
\usepackage[pdftex]{graphicx}
\usepackage{algorithmic}
\usepackage{amsmath}
\usepackage{amssymb}
\usepackage{multirow}
\usepackage{color}
\usepackage{longtable}
\usepackage{array}
\usepackage{url}
\usepackage[ruled,linesnumbered]{algorithm2e}
\usepackage{enumerate}
\usepackage{float}
\usepackage{mathtools}
\usepackage{bbm}
\usepackage{cite}
\usepackage{comment}
\newtheorem{theorem}{Theorem}[section]
\newtheorem{definition}[theorem]{Definition}
\newtheorem{lemma}[theorem]{Lemma}
\newtheorem{example}[theorem]{Example}
\newtheorem{construction}[theorem]{Construction}
\newtheorem{corollary}[theorem]{Corollary}
\newtheorem{remark}[theorem]{Remark}
\newtheorem{observation}[theorem]{Observation}

\begin{document}

\title{Improved constructions of permutation and multi-permutation codes correcting a burst of stable deletions}

\author{Yubo~Sun, Yiwei~Zhang, and Gennian~Ge%
\thanks{The research of G. Ge was supported by the National Key Research and Development Program of China under Grant 2020YFA0712100 and Grant 2018YFA0704703, the National Natural Science Foundation of China under Grant 11971325 and Grant 12231014, and Beijing Scholars Program.}
\thanks{The research of Y. Zhang was supported by the National Key Research and Development Program of China under Grant 2022YFA1004900 and Grant 2021YFA1001000, the National Natural Science Foundation of China under Grant 12001323, and by the Shandong Provincial Natural Science Foundation under Grant ZR2021YQ46.}
\thanks{Y. Sun and G. Ge are with the School of Mathematical Sciences, Capital Normal University, Beijing 100048, China (e-mail: 2200502135@cnu.edu.cn, gnge@zju.edu.cn).}
\thanks{Y. Zhang is with the Key Laboratory of Cryptologic Technology and Information Security, Ministry of Education, and also with the School of Cyber Science and Technology, Shandong University, Qingdao, Shandong 266237, China (e-mail: ywzhang@sdu.edu.cn).}
}

\maketitle

\begin{abstract}

Permutation codes and multi-permutation codes have been widely considered due to their various applications, especially in flash memory. In this paper, we consider permutation codes and multi-permutation codes against a burst of stable deletions.
In particular, we propose a construction of permutation codes correcting a burst stable deletion of length $s$, with redundancy $\log  n+ 2\log \log  n+O(1)$.
Compared to the previous known results, our improvement relies on a different strategy to retrieve the missing symbol on the first row of the array representation of a permutation. We also generalize our constructions for multi-permutations and the variable length burst model.
Furthermore, we propose a linear-time encoder with optimal redundancy for single stable deletion correcting permutation codes.
\end{abstract}


\section{Introduction}

Flash memory is now widely used, especially in portable storage devices, due to its physical reliability, high storage density, and relatively low cost. Coding schemes for error-correction in flash memory have received considerable attention in recent years. In particular, the {\it rank modulation scheme}, which uses permutations to encode messages, has been proposed by Jiang et al. \cite{Jiang-09-IT-RM} to cope with errors caused by charge leakage or charge overshooting during programming. Flash memories are comprised of blocks of cells, and in the rank modulation scheme information is characterized by the relative charge order among a group of cells instead of their absolute charge levels.
In this setup, each group of cells induces a permutation.
Therefore, the research on permutation codes under different metrics is one of the main topics in the rank modulation scheme of flash memories \cite{Jiang-10-IT-RM,Barg-10-IT-RM,Farnoud-13-IT-FM_UM,Zhou-15-IT-RM,Gad-12-ISIT-FM,Horovitz-19-IT-RM}. Afterwards, multi-permutation codes, as a generalization of permutation codes with more flexibility, are also introduced in the rank modulation scheme of flash memories \cite{Gad-12-ISIT-FM,Sala-14-ISIT-MPD,Sala-13-Commun-NVM,Hassanzadeh-14-JSAC-MP_UM}.

When reading and comparing the charge levels of the cells, some cells might be corrupted and the relative order of charge levels cannot be read correctly. This leads to either erasures or deletions in the permutation, depending on whether the erroneous coordinates are known or unknown. Gabrys et al. \cite{Gabrys-16-IT-PD} formally proposed the models of stable/unstable erasures/deletions in permutation codes for flash memories. Here the term stable, also known as symbol-invariant, refers to the case that after deletion/erasure the remaining symbols stay invariant. The term unstable, also known as permutation-invariant, refers to the case that after deletion/erasure the remaining symbols are sorted again, i.e., a new permutation is formed based on the relative orders among the remaining cells. For example, consider a permutation $(2,3,1,5,4)$ and assume one erasure/deletion error occurs on the second coordinate. A stable erasure will result in $(2,?,1,5,4)$ and an unstable erasure will result in $(2,?,1,4,3)$. Similarly, a stable deletion will result in $(2,1,5,4)$ while an unstable deletion will result in $(2,1,4,3)$. See later for a formal definition. In particular, the unstable erasure model is a quite common event in flash memory and it was shown to be equivalent to the stable deletion model \cite{Gabrys-16-IT-PD}. As the stable deletion model is simpler in its mathematical formulations, more subsequent works were focused on the stable deletion model rather than the unstable erasure model. Afterwards, these concepts were further extended to multi-permutation codes by Sala et al. \cite{Sala-14-ISIT-MPD}.

For single deletion in permutation codes, Levenshtein \cite{Levenshtein-92-DM-ID} showed that a set of permutations whose signatures form a binary VT code is capable of correcting a single stable deletion.
Gabrys et al. \cite{Gabrys-16-IT-PD} also used binary VT codes to design asymptotically optimal single unstable deletion correcting permutation codes.
For multiple deletions in permutation codes, researchers focus on burst deletions. One of the most important reasons is that deletions tend to occur in consecutive cells in flash memories due to capacitative coupling \cite{Lee-02-NAND,Prall-07-FM}.
In fact, not only in flash memories, but also in DNA storage \cite{Schoeny-17-IT-BD,Gabrys-18-IT-D_T} and racetrack memories \cite{Chee-18-IT-RM}, errors manifest themselves in burst and tend to cluster together.
Note that the term of burst error is slightly different between substitution error models and deletion error models. To avoid confusion, we clarify the term of burst deletion as follows. Given an integer $s$,
\begin{itemize}
  \item a burst of consecutive deletions of length $s$ \cite{Levenshtein-70-STR-BD,Schoeny-17-IT-BD,Saeki-18-ISIT-BD,Chee-20-IT-PBSD}, refers to the error where exactly $s$ consecutive symbols are deleted;
  \item a burst of consecutive deletions of length at most $s$ \cite{Schoeny-17-IT-BD,Lenz-20-ISIT-BD,Chee-20-IT-PBSD}, refers to the error where at most $s$ consecutive symbols are deleted;
  \item a burst of nonconsecutive deletions of length at most $s$ \cite{Schoeny-17-IT-BD}, also known as the localized deletions \cite{Bitar-21-ISIT-LD,Hanna-21-IT-LD}, refers to the error where some symbols are deleted and the deleted coordinates are within a block of length at most $s$.
\end{itemize}

In this paper we only consider the first two types of errors, where the first is called a fixed length burst and the second is called a variable length burst.
A series of works \cite{Chee-15-ISIT-PBUD,Chee-17-ISIT-PBSD,Chee-20-IT-PBSD,Han-18-CL-PBSD,Han-20-CL-PBSD} have considered permutation codes and multi-permutation codes against burst deletions, for both stable and unstable types of errors and for both fixed length burst and variable length burst models.
Among all the known results, so far the best permutation codes in $S_n$ correcting a fixed length burst of stable deletions have redundancy $2\log  n +O(1)$, while the lower bound of the redundancy for such codes derived by a sphere-packing argument is only $\log  n$ \cite{Chee-20-IT-PBSD}. In this paper, we focus on burst stable deletions and one of our main results is to narrow the previous gap by constructing a code with redundancy $\log  n + 2\log \log  n +O(1)$.
The main challenge in this area is to retrieve the missing symbol on the first row of the array representation of a permutation. Since the alphabet size is $n$, the traditional method which simply applies a sum constraint modulo $n$ on the first row will require $\log n$ bits of redundancy. One of our main contributions is to provide a novel strategy to retrieve the missing symbol with less redundancy. Using this strategy, we improve the redundancy of burst stable deletion correcting permutation codes. We then generalize our construction of permutation codes against fixed length burst stable deletions into variable length burst, and also their multi-permutation counterparts.
Finally, we investigate the coding scheme in permutation codes and present a linear-time encoder correcting a single stable deletion with $\log n$ bits of redundancy.

The rest of this paper is organized as follows.
Section \ref{Sec:pre} introduces the relevant definitions and notations used throughout the paper as well as the previous related results.
In Section \ref{Sec:symbol} we present a novel approach to retrieve the missing symbol on the first row of the array representation of a permutation or a multi-permutation, which plays a key role in our main constructions.
In Section \ref{Sec:fixed} and Section \ref{Sec:variable}, we study permutation and multi-permutation codes against burst stable deletions, for fixed length burst and variable length burst, respectively.
In Section \ref{Sec:Enc}, we present a linear-time encoder for single stable deletion correcting permutation codes.
Finally Section \ref{Sec:concl} concludes the paper.

\section{Preliminaries} \label{Sec:pre}

We begin with introducing the relevant definitions and notations used throughout the paper. Let $\Sigma_q$ denote the alphabet $\{ 0,1, \ldots, q-1 \}$ with $q$ symbols and $\Sigma_q^n$ denote the set of all sequences of length $n$ over $\Sigma_q$. Similarly, $\mathbb{Z}^n$ denotes the set of vectors of length $n$ over integers.
A vector is always denoted by bold letters, such as $\boldsymbol{x}$, and its $i$-th entry is denoted as $x_i$. For two positive integers $i,j$ with $i \leq j$, the set of integers $\{i,i+1,\dots,j\}$ is denoted as $[i,j]$ and the set $[1,j]=\{1,2,\ldots,j \}$ is abbreviated as $[j]$. For two vectors $\boldsymbol{u} = \left( u_1,u_2,\ldots,u_t \right)$ and $\boldsymbol{v} = \left( v_1,v_2,\ldots,v_n \right)$, let $(\boldsymbol{u},\boldsymbol{v})$ denote their \emph{concatenation} $\left( u_1,u_2,\ldots,u_t,v_1,v_2,\ldots,v_n \right)$. If $t\leq n$ and there exist some integers $1 \leq i_1 < i_2 < \cdots < i_t \leq n$ such that $u_j= v_{i_j}$ for all $1 \leq j \leq t$, then we say that $\boldsymbol{u}$ is a \emph{subsequence} of $\boldsymbol{v}$. Furthermore, if $i_{j+1} = i_j + 1$ for $1 \leq j \leq t-1$, then $\boldsymbol{u}$ is called a \emph{consecutive subsequence} of $\boldsymbol{v}$.

Let $S_n$ be the symmetric group of order $n$.
For any permutation $\boldsymbol{\sigma} = \left( \sigma_1,\sigma_2,\ldots,\sigma_n \right) \in S_n$, define $\boldsymbol{\sigma}(\mathcal{I}) = \{\sigma_i : i \in \mathcal{I}\}$ for any subset $\mathcal{I} \subseteq [n]$. For any subset $X$ of positive integers and a positive integer $k\notin X$, define $k(X) = k - |\{i \in X : i \leq k\}|$. Thus $k(X)=k'$ means that the positive integer $k$ becomes the $k'$-th smallest positive integer in $\mathbb{Z}^{+}\setminus X$. For example, if $X=\{2,4,6\}$ then $1(X)=1, 3(X)=2, 5(X)=3$, and $k(X)=k-3$ for $k\ge 7$.

As with most existing works concerning burst deletions, it is convenient to represent a vector in the form of an array. Given an integer $s$, we always assume that $n=st$ holds for some integer $t$. A vector $\boldsymbol{\sigma} = (\sigma_1,\sigma_2,\ldots,\sigma_n)$ is represented as the following $s\times t$ array
\begin{equation*}
    \begin{pmatrix}
        \sigma_1 & \sigma_{s+1} & \cdots & \sigma_{(t-1)s+1} \\
        \sigma_2 & \sigma_{s+2} & \cdots & \sigma_{(t-1)s+2} \\
        \vdots   & \vdots       & \ddots & \vdots            \\
        \sigma_s & \sigma_{2s}  & \cdots & \sigma_{n}
    \end{pmatrix},
\end{equation*}
where the $i$-th row and $j$-th column of the array is denoted as $\boldsymbol{\sigma}_{(s,i)}$ and $\boldsymbol{\sigma}^{(s,j)}$, respectively.

Now we formally define the types of errors and the corresponding codes studied in this paper.

\subsection{Permutation codes}

\begin{definition}[Stable deletions in permutations]\label{def:BSD}
Let $\boldsymbol{\sigma} = \left( \sigma_1,\sigma_2,\ldots,\sigma_n \right)$ be a permutation in $S_n$ and $\mathcal{I}$ be a subset of $[n]$ of size $s$. We say that $\boldsymbol{\sigma}$ suffers $s$ \emph{stable deletions} in $\mathcal{I}$ and results in $\boldsymbol{\sigma}' = \left( \sigma_1',\sigma_2',\ldots,\sigma_{n-s}' \right)$, if for all $k \in [1,n] \setminus \mathcal{I}$ and $i = k(\mathcal{I})$, we have $\sigma_{i}' = \sigma_k$.
\end{definition}

\begin{definition}[Unstable deletions in permutations]
Let $\boldsymbol{\sigma} = \left( \sigma_1,\sigma_2,\ldots,\sigma_n \right)$ be a permutation in $S_n$ and $\mathcal{I}$ be a subset of $[n]$ of size $s$. We say that $\boldsymbol{\sigma}$ suffers $s$ \emph{unstable deletions} in $\mathcal{I}$ and results in $\boldsymbol{\sigma}' = \left( \sigma_1',\sigma_2',\ldots,\sigma_{n-s}' \right)$, if for all $k \in [1,n] \setminus \mathcal{I}$ and $i = k(\mathcal{I})$, we have $\sigma_{i}' = \sigma_k(\boldsymbol{\sigma}(\mathcal{I}))$.
\end{definition}

\begin{example}
Let $\boldsymbol{\sigma} = \left(2,4,5,1,6,3 \right) \in S_6$ and let the set of deleted coordinates be $\mathcal{I} = \{2,4\}$.
In the model of stable deletions, the resultant vector is $(2,5,6,3)$. In the model of unstable deletions, the resultant vector is $(1,3,4,2)$.
\end{example}

Note that in the model of stable deletions the remaining symbols stay invariant and thus the resultant vector may no longer be a permutation, whereas in the model of unstable deletions the remaining symbols may change and form a new permutation in $S_{n-s}$ based on their relative orders. This is also why the two terms were named as symbol-invariant and permutation-invariant in \cite{Gabrys-16-IT-PD}. In this paper we only consider stable deletions.

\begin{definition}[Burst stable deletions in permutations]
In Definition \ref{def:BSD}, if $\mathcal{I}$ is an interval of length $s$ of the form $\mathcal{I} = [i,i+s-1]$ for some $i \in [n-s+1]$, then we say that $\boldsymbol{\sigma}$ suffers a \emph{burst stable deletion} of length $s$.
\end{definition}

Immediately we have the following observation.

\begin{observation}\label{obs:span}
In the $s\times t$ array representation of a permutation  $\boldsymbol{\sigma} \in S_n$, a burst deletion of length $s$ results in exactly one deletion on each row and the deleted coordinates are within at most two adjacent columns.
\end{observation}

\begin{definition}\label{def:BSDC}
Let $\mathcal{B}_s(\boldsymbol{\sigma})$ be the set of all subsequences which are resulted from $\boldsymbol{\sigma}$ after $s$ stable deletions. A code $\mathcal{C} \subseteq S_n$ is called an $s$-\emph{SD permutation code}, if for any two distinct permutations $\boldsymbol\sigma_1\in \mathcal{C},\boldsymbol\sigma_2 \in \mathcal{C}$, it holds that $\mathcal{B}_s(\boldsymbol{\sigma}_1) \cap \mathcal{B}_s(\boldsymbol{\sigma}_2) = \varnothing$. That is, $\mathcal{C}$ can correct exactly $s$ stable deletions.
Similarly we can define the following classes of codes:
\begin{itemize}
  \item An $s$-\emph{BSD permutation code} can correct a burst stable deletion of length $s$.
  \item An $^\leq s$-\emph{BSD permutation code} can correct a burst stable deletion of length at most $s$.
  %\item An $s$-\emph{UD permutation code} can correct exactly $s$ unstable deletions.
\end{itemize}
\end{definition}

\subsection{Multi-permutation codes}

A multi-permutation is a generalization of a permutation where each element appears multiple times that are known a priori.

\begin{definition}\label{def:MP}
Let $n, w$ be positive integers such that $w\le n$. A vector $\boldsymbol{r} = \left( r_1,r_2,\ldots,r_w \right)\in \mathbb{Z}^w$ is called a \emph{multiplicity vector} if $n = \sum_{i=1}^w r_i$ and $r_i \geq 1$ for $1\leq i \leq w$. Let $M(n, \boldsymbol{r})$ denote the multiset
\begin{equation*}
\{\underbrace{1,\ldots,1}_{r_1},\underbrace{2,\ldots,2}_{r_2},\ldots,\underbrace{w,\ldots,w}_{r_w}\}.
\end{equation*}
A \emph{multi-permutation} $\boldsymbol{\sigma} = \left( \sigma_1,\sigma_2,\ldots,\sigma_n \right)$ is an arrangement of the elements of the multiset $M(n,\boldsymbol{r})$, i.e., the number $i$ appears exactly $r_i$ times in $\boldsymbol{\sigma}$, $1 \leq i \leq w$. Let $S_n^{\boldsymbol{r}}$ denote the set of all multi-permutations on $M(n, \boldsymbol{r})$. In particular, if $n=wr$ and $r_1 = r_2 = \cdots = r_w = r$, we say that the multiset and the corresponding multi-permutations are \emph{regular}.
\end{definition}

Terminologies regarding permutations can be naturally generalized for multi-permutations. For the interest of this paper we only list the following definitions.

\begin{definition}
Let $\boldsymbol{r} = \left( r_1,r_2,\ldots,r_w \right)$ be a multiplicity vector. Let $\boldsymbol{\sigma} = \left( \sigma_1, \sigma_2,\ldots,\sigma_n \right)$ be a multi-permutation in $S_n^{\boldsymbol{r}}$ and let $\mathcal{I} = [i,i + s-1]$ be an interval of length $s$ for some $i \in [n-s+1]$. We say that $\boldsymbol{\sigma}$ suffers a \emph{burst stable deletion} of length $s$ in $\mathcal{I}$, if the resultant vector is $\left( \sigma_1,\ldots,\sigma_{i-1}, \sigma_{i+s},\ldots,\sigma_n \right)$.
\end{definition}

\begin{example}
Let $\boldsymbol{\sigma} = \left(3, 1, 3, 2, 2, 1, 2, 1, 3\right) \in S_9^{(3,3,3)}$ and $\mathcal{I} = [2,4]$. If $\boldsymbol{\sigma}$ suffers a burst stable deletion of length $3$ in $\mathcal{I}$, the resultant vector is $\left(3, 2, 1, 2, 1, 3\right)$.
\end{example}

\begin{definition}
Let $\boldsymbol{r}$ be a multiplicity vector. We say a code $\mathcal{C} \subseteq S_n^{\boldsymbol{r}}$ is an $s$-\emph{BSD multi-permutation code} if it can correct a burst stable deletion of length $s$, or an $^\leq s$-\emph{BSD multi-permutation code} if it can correct a burst stable deletion of length at most $s$.
\end{definition}

\subsection{$q$-ary codes}

For $q$-ary codes the deletion errors are naturally stable deletions.

\begin{definition}
Let $\boldsymbol{x} = \left( x_1, x_2,\ldots,x_n \right)$ be a sequence in $\Sigma_q^n$ and let $\mathcal{I} = [i,i + s-1]$ be an interval of length $s$ for some $i \in [n-s+1]$. We say that $\boldsymbol{x}$ suffers a \emph{burst stable deletion} of length $s$ in $\mathcal{I}$, if the resultant vector is $\left( x_1,\ldots,x_{i-1}, x_{i+s},\ldots,x_n \right)$.
\end{definition}

\begin{definition}
A code $\mathcal{C} \subseteq \Sigma_q^n$ is a \emph{$q$-ary $s$-burst deletion correcting code} if it can correct a burst stable deletion of length $s$, or a \emph{$q$-ary $^\leq s$-burst deletion correcting code} if it can correct a burst stable deletion of length at most $s$.
\end{definition}

\subsection{Previous results}

In this subsection we present some known constructions of codes against deletion errors or burst deletion errors.
The history of codes against deletion errors dates back to 1966, when Levenshtein \cite{Levenshtein-66-SPD-1D} showed that the celebrated {\it Varshamov-Tenengol'ts (VT) codes} \cite{VTcode-AT65} are binary codes capable of correcting one deletion error. Recently, due to various applications in synchronization error-correction \cite{Mitzenmacher-09-PS-syn,Mercier-10-CST-syn,Haeupler-17-ACM-syn}, DNA storage \cite{Yazdi-15-TMBMC-DNA,Gabrys-18-IT-D_T,Levy-19-IT-MUC}, and racetrack memories \cite{Chee-18-IT-RM} et al., much progress has been made on binary or $q$-ary codes against multiple deletion errors \cite{Tenengol'ts-84-IT-q_D,Gabrys-19-IT-2D,Sima-20-IT-2D,Guruswami-21-IT-2D,Sima-21-IT-kD,Brakensiek-18-IT-kD,Sima-20-ISIT-tD,Sima-20-ISIT-q-tD,Hanna-19-IT-kD}.

To evaluate a code $\mathcal{C}\subseteq \mathcal{S}$, where the underlying set $\mathcal{S}$ can be $\Sigma_2^n$, $S_n$, or $S_n^{\boldsymbol{r}}$, we either calculate its \emph{size} $|\mathcal{C}|$ or analyze its \emph{redundancy}, defined as $\log |\mathcal{S}|-\log|\mathcal{C}|$, where the logarithm base is $2$.

The following list is not intended as a full survey of all the known results, but only focuses on the more related codes which will be the building blocks in our main construction. Related important terminologies and ideas are attached to specific constructions.

\subsubsection{Single deletion correcting code}

\begin{itemize}
  \item {\it Binary single deletion correcting code}:  For $a \in \mathbb{Z}_{n+1}$, the \emph{VT code} $\mathrm{VT}_a(n)$ is defined as
      \begin{equation*}
        \mathrm{VT}_a(n) = \left\{ \boldsymbol{x} \in \Sigma_2^n : \mathrm{VT}(\boldsymbol{x}) \equiv a \mod (n+1)\right\},
      \end{equation*}
    where $\mathrm{VT}(\boldsymbol{x})= \sum_{i=1}^n i x_i$ is called the \emph{VT syndrome} of $\boldsymbol{x}$.
    Levenshtein \cite{Levenshtein-66-SPD-1D} showed that $\mathrm{VT}_a(n)$ is a binary single deletion correcting code.

  \item {\it $q$-ary single deletion correcting code}: For any sequence $\boldsymbol{x} = \left( x_1,x_2,\ldots,x_n \right)\in \Sigma_q^n$, we define  $\alpha(\boldsymbol{x}) = (\alpha(x_1),\alpha(x_2),\ldots,\alpha(x_{n-1}))$, with $\alpha(x_i)=1$ if $x_{i+1} \geq x_i$ and $\alpha(x_i)=0$ otherwise, for all $i \in [n - 1]$. The vector $\alpha(\boldsymbol{x})$ represents the relative order between consecutive symbols and is called the \emph{signature} of $\boldsymbol{x}$. For $a \in \mathbb{Z}_n$ and $b \in \mathbb{Z}_q$, define
    \begin{align*}
        \mathrm{VT}_{a,b}(n,q) = \big\{\boldsymbol{x} \in \Sigma_q^n :
        &~ \alpha(\boldsymbol{x}) \in VT_a(n - 1), \\
        &~ \mathrm{Sum}(\boldsymbol{x}) \equiv b \pmod{q} \big\},
    \end{align*}
    where $\mathrm{Sum}(\boldsymbol{x})= \sum_{i=1}^n x_i$. Tenengol'ts \cite{Tenengol'ts-84-IT-q_D} showed that $\mathrm{VT}_{a,b}(n,q)$ is a $q$-ary single deletion correcting code.

  \item {\it Single stable deletion correcting permutation code}: Levenshtein \cite{Levenshtein-92-DM-ID} showed that permutations whose signature satisfies a VT constraint form a single stable deletion correcting permutation code. Note that such a code can be seen as a subcode of $\mathrm{VT}_{a,b}(n,q)$ restricting to $S_n$, with $q=n$ and $b=1+2+\dots+n$.
\end{itemize}

\subsubsection{$P$-bounded single deletion correcting code (correcting a deletion given the additional knowledge of $P$ consecutive coordinates which contain the erroneous coordinate)}

\begin{itemize}
    \item {\it Binary $P$-bounded single deletion correcting code}: For $a \in \mathbb{Z}_P$ and $b \in \mathbb{Z}_2$, the binary \emph{shifted VT code} (SVT code) $\mathrm{SVT}_{a,b}(n,P)$ is defined as
    \begin{align*}
        \mathrm{SVT}_{a,b}(n,P) = \big\{ \boldsymbol{x} \in \Sigma_2^n :
        &~ \mathrm{VT}(\boldsymbol{x}) = a \pmod{P}, \\
        &~ \mathrm{Sum}(\boldsymbol{x})= b \pmod{2} \big\}.
    \end{align*}
    Schoeny et al. \cite{Schoeny-17-IT-BD} showed that $\mathrm{SVT}_{a,b}(n,P)$ is a binary $P$-bounded single deletion correcting code. Note that compared to binary VT codes, in binary SVT codes the VT syndrome is computed modulo $P$ instead of $n+1$.

    \item {\it $q$-ary $P$-bounded single deletion correcting code}: For $a \in \mathbb{Z}_P$, $b \in \mathbb{Z}_2$, and $c \in \mathbb{Z}_q$, the $q$-ary SVT code $\mathrm{SVT}_{a,b,c}(n,P,q)$ is defined as:
    \begin{align*}
        \mathrm{SVT}_{a,b,c}(n,P,q) = \big\{ \boldsymbol{x} \in \Sigma_q^n :  \mathrm{Sum}(\boldsymbol{x})= c \pmod{q}, \\
         \alpha(\boldsymbol{x}) \in \mathrm{SVT}_{a,b}(n-1,P) \big\}.
    \end{align*}
    Schoeny et al. \cite{Schoeny-17-SVT} showed that $\mathrm{SVT}_{a,b,c}(n,P,q)$ is an $q$-ary $P$-bounded single deletion correcting code. Compared to $q$-ary VT codes, in $q$-ary SVT codes the signature of each codeword lies in a binary SVT code instead of a binary VT code.
\end{itemize}

\subsubsection{$s$-burst deletion correcting code}

\begin{itemize}
  \item {\it Binary $s$-burst deletion correcting code}: Schoeny et al. \cite{Schoeny-17-IT-BD} proposed a construction of binary $s$-burst deletion correcting codes
  as follows. They represented each codeword in the form of an $s \times t$ array and constructed the codes for each row independently. The first row is a subcode of the VT code satisfying additional constraints known as the \emph{run-length limited constraints}. A \emph{run} in a sequence is a maximal consecutive subsequence consisting of the same symbol and a run-length limited constraint requires that the length of the longest run in a codeword is upper bounded by a predetermined value $P$. Under such constraints, by decoding the first row one can locate the deleted coordinate within a bounded interval of length $P$ and thus provide additional information for the decoding of the remaining rows. Each of the remaining rows applies a $(P+1)$-bounded single deletion correcting code. Here $P$ is set to be of order $\log  {n}$ so as to reduce the overall redundancy.

  \item {\it $q$-ary $s$-burst deletion correcting code}: Saeki et al. \cite{Saeki-18-ISIT-BD} generalized the binary codes in \cite{Schoeny-17-IT-BD} by substituting the binary VT or SVT accessories therein as their $q$-ary versions, leading to a $q$-ary $s$-burst deletion correcting code.

  \item {\it $s$-burst stable deletion correcting permutation code and multi-permutation code}: Han et al. \cite{Han-18-CL-PBSD,Han-20-CL-PBSD} used an interleaving technique to construct $s$-burst stable deletion correcting permutation codes and multi-permutation codes. The current best result comes from Chee et al. \cite{Chee-20-IT-PBSD} with redundancy $2\log  n+O(1)$. They further generalized their construction of permutation codes to multi-permutation codes. Details of their construction are postponed to the next section, so as to bring out our main idea for improvements.
\end{itemize}

\subsubsection{$^{\leq}s$-burst deletion correcting code}

\begin{itemize}
    \item {\it Binary $^\leq s$-burst deletion correcting code}: Lenz et al. \cite{Lenz-20-ISIT-BD} followed the above framework of Schoeny et al. but encoded each row with a different strategy. Their main tool is an additional $(\boldsymbol{p},\delta)$-\emph{dense constraint} which requires that any consecutive subsequence of length $\delta$ contains at least one pattern $\boldsymbol{p}$ as a consecutive subsequence. A burst deletion may destroy or generate the occurrences of $\boldsymbol{p}$ and thus by observing the pattern one can locate the burst error within a bounded interval. With this additional knowledge, several binary SVT codes are then applied for correction.
    \item {\it $^{\leq}s$-burst deletion correcting permutation code}: Chee et al. \cite{Chee-20-IT-PBSD} constructed $^{\leq}s$-burst deletion correcting permutation codes by taking the intersection of their $s'$-burst stable deletion correcting permutation codes for all $1 \leq s' \leq s$, with some slight modifications.
\end{itemize}

\section{Retrieve the missing symbol in the first row of the array representation} \label{Sec:symbol}

By viewing a permutation as an $s\times \frac{n}{s}$ array, a burst stable deletion of length $s$ results in exactly one stable deletion on each row. We briefly go over the construction of Chee et al. \cite{Chee-20-IT-PBSD}. They imposed an $n$-ary VT constraint to the first row of the array, which would help retrieve the missing symbol in the first row of the array representation. As for permutations, this means the erroneous coordinate in the first row can be exactly located. Then the erroneous coordinates in the other rows are within two adjacent columns. Next, they recorded two values which add up the permutation rank (to be defined later in Definition \ref{def:permutation rank}) of the vector concatenated by two consecutive columns, where the two sums begin with the first and the second column, respectively. One of these two sums will contain the concatenation of the two erroneous columns and reveals its permutation rank. With the permutation rank and the known set of missing symbols, one can perform the correction.

Most redundancy ($2\log  n$ bits) of the codes above come from the $n$-ary VT constraint in the first row. The $n$-ary VT constraint contains two parts, the binary VT constraint on the signature and a sum constraint of all entries modulo $n$. While the first part seems inevitable, the second part can be improved, in the sense that the large alphabet size $n$ can be significantly reduced if we have some additional constraint on the error pattern. Our key idea for improvements is to retrieve the missing symbol in the first row in a different way with less redundancy, consisting of the following three steps. For the rest of this paper we fix $P=\lceil \log {\frac{4n}{s}} \rceil$ and assume $n$ is a multiple of $2Ps$.

$\bullet$ Step 1. Locate the erroneous coordinate of the first row of the array within an interval of length $P$.

$\bullet$ Step 2. Find a proper block of $2P$ consecutive columns which contains all the erroneous coordinates.

$\bullet$ Step 3. Retrieve the missing symbol of the first row, by analyzing the $2P$ consecutive columns from the previous step.

Step 1 can be guaranteed by applying some additional constraint on the first row. Since the techniques for Step 1 vary for permutation/multi-permutation codes against fixed/variable length burst, we leave it later to the next two sections when we introduce each specific code. In this section, we assume that we have already finished Step 1 and know the interval of length $P$ containing the erroneous coordinate of the first row, and explain Steps 2-3 (which are two universal steps regardless of permutations/multi-permutations and fixed/variable burst stable deletions).

\subsection{Step 2: Find $2P$ consecutive columns which contain all erroneous coordinates}

\begin{definition}\label{def:partition}
For appropriate integers $s$, $P$, and $i$. Let $\boldsymbol{\sigma}(s,P,i) \triangleq (\boldsymbol{\sigma}^{(s,(i-1)P+1)}, \boldsymbol{\sigma}^{(s,(i-1)P+2)}, \ldots, \boldsymbol{\sigma}^{(s,iP)})$, be the subarray of $\boldsymbol{\sigma}$ consisting of the columns indexed from $(i-1)P+1$ to $iP$, i.e.
\begin{align*}
    &\boldsymbol{\sigma}(s,P,i)= \\
    &\begin{pmatrix}
        \sigma_{(i-1)Ps+1} & \sigma_{((i-1)P+1)s+1} & \cdots & \sigma_{(iP-1)s+1}\\
        \sigma_{(i-1)Ps+2} & \sigma_{((i-1)P+1)s+2} & \cdots & \sigma_{(iP-1)s+2}\\
        \vdots & \vdots & \ddots & \vdots \\
        \sigma_{((i-1)P+1)s} & \sigma_{((i-1)P+2)s} & \cdots & \sigma_{iPs}\\
    \end{pmatrix}.
\end{align*}
\end{definition}

Suppose the erroneous coordinate of the first row is in the interval $\mathcal{I}= [i,j]$ with $j-i+1 \leq P$. Then the erroneous coordinates in the other rows are within the columns indexed by $[i-1,j]$ if $i \geq 2$, or $[1,j]$ if $i=1$. According to Definition \ref{def:partition}, there is at least one block $\boldsymbol{\varsigma}= (\boldsymbol{\sigma}(s,P,k),\boldsymbol{\sigma}(s,P,k+1))$ containing the erroneous columns, as shown in Figure \ref{fig:block}.

\begin{figure*}[t]
\begin{gather*}
    \boldsymbol{\sigma}=
    \begin{pmatrix}
        \sigma_1 & \cdots & \sigma_{(k-1)Ps+1} & \cdots & \sigma_{(i-2)s+1} & \cdots & \sigma_{(j-1)s+1} & \cdots & \sigma_{((k+1)P-1)s+1} & \cdots & \sigma_{(t-1)s+1} \\
        \vdots & \ddots & \vdots & \ddots & \vdots & \ddots & \vdots & \ddots & \vdots & \ddots & \vdots \\
        \sigma_s & \cdots & \sigma_{((k-1)P+1)s} & \cdots & \sigma_{(i-1)s} & \cdots & \sigma_{js} & \cdots & \sigma_{(k+1)Ps} & \cdots & \sigma_{n} \\
    \end{pmatrix} \\
    \downarrow \\
    \boldsymbol{\varsigma}= (\boldsymbol{\sigma}(s,P,k),\boldsymbol{\sigma}(s,P,k+1))=
    \begin{pmatrix}
        \sigma_{(k-1)Ps+1} & \cdots & \sigma_{(i-2)s+1} & \cdots & \sigma_{(j-1)s+1} & \cdots & \sigma_{((k+1)P-1)s+1}\\
        \vdots & \ddots & \vdots & \ddots & \vdots & \ddots & \vdots \\
        \sigma_{((k-1)P+1)s} & \cdots & \sigma_{(i-1)s} & \cdots & \sigma_{js} & \cdots & \sigma_{(k+1)Ps}\\
    \end{pmatrix}
\end{gather*}
\caption{Find $2P$ consecutive columns which contain all erroneous coordinates.}
\label{fig:block}
\end{figure*}

Moreover, all entries in the block $\boldsymbol{\varsigma}$ are known (since it is the complement of the set of entries on the other non-erroneous columns).

\subsection{Step 3: Retrieve the missing symbol of the first row}

Clearly, to retrieve the missing symbol of the first row, one may simply apply a sum constraint modulo $n$ for the first row $\boldsymbol{\sigma}_{(s,1)}$. However, it would still require $\log  n$ bits of redundancy. Below we will show that, with the help of Step 2, the missing symbol of the first row can be retrieved with less redundancy.  We need the following definition.

\begin{definition}\label{def:induced}
Let $\boldsymbol{u} = (u_1,u_2,\ldots,u_n)$ be a vector with integer entries. Define $\beta(\boldsymbol{u})=(\beta(\boldsymbol{u})_1,\beta(\boldsymbol{u})_2,\dots,\beta(\boldsymbol{u})_n)\in S_n$ to be the permutation induced by $\boldsymbol{u}$ as follows:
\begin{align*}
    \beta(\boldsymbol{u})_i
    & = | \left\{j : u_j < u_i,1 \leq j \leq n \right\}| \\
    & \quad \quad \quad + | \left\{j : u_j = u_i,1 \leq j \leq i \right\}|.
\end{align*}
That is, we rank the entries of $\boldsymbol{u}$ in an increasing order and the repeated symbols are ordered according to their appearances. $\beta(\boldsymbol{u})$ is called the \emph{permutation projection} of $\boldsymbol{u}$ and $\beta(\boldsymbol{u})_i$ is the \emph{order} of the symbol $u_i$ in $\boldsymbol{u}$.
\end{definition}

\begin{example}
Let $\boldsymbol{u} = (6,2,5,1,8,4), \boldsymbol{v} = (1, 1, 2, 2, 1, 2)$. Then $\beta(\boldsymbol{u}) =(5,2,4,1,6,3)$ and $\beta(\boldsymbol{v})=(1, 2, 4, 5, 3, 6)$.
\end{example}

\begin{lemma}\label{lem:symbol}
Let $\mathcal{J}=\{\varsigma_1,\varsigma_2,\ldots,\varsigma_{2Ps}\}\subseteq [n]$ be a set of $2Ps$ distinct elements and $\boldsymbol{\varsigma}=(\varsigma_1,\varsigma_2,\ldots,\varsigma_{2Ps})$ be a vector written in an $s \times 2P$ array. Suppose we know $\mathrm{Sum}(\beta(\boldsymbol{\varsigma})_{(s,1)}) \equiv c \pmod{2Ps}$ with $c \in \mathbb{Z}_{2Ps}$. If $\boldsymbol{\varsigma}$ suffers a burst stable deletion of length $s$, then we can retrieve the missing symbol of the first row given the knowledge of the set $\mathcal{J}$.
\end{lemma}

\begin{IEEEproof}
Assume $\boldsymbol{\varsigma}$ suffers a burst stable deletion of length $s$ and the resultant vector is $\boldsymbol{v}= (v_1,v_2,\ldots,v_{(2P-1)s})$.
Since the entries of $\boldsymbol{\varsigma}$ are distinct, its permutation projection naturally induces a bijection $f_\beta: \{\varsigma_{1}, \varsigma_2, \ldots, \varsigma_{2Ps}\} \rightarrow [2Ps]$ where $f_\beta(\varsigma_{i})=\beta(\boldsymbol{\varsigma})_i$. Let $\tilde{\boldsymbol{v}}= (f_\beta(v_1),f_\beta(v_2),\ldots,f_\beta(v_{(2P-1)s}))$. Then $\tilde{\boldsymbol{v}}$ can be seen as a resultant vector of $\beta(\boldsymbol{\varsigma})$ suffering a burst stable deletion of length $s$. Consider their first rows, $\tilde{\boldsymbol{v}}_{(s,1)} \in \mathcal{B}_1(\beta(\boldsymbol{\varsigma})_{(s,1)})$. Therefore, as long as we know that $\mathrm{Sum}(\beta(\boldsymbol{\varsigma})_{(s,1)}) \equiv c \pmod{2Ps}$, the missing symbol of $\boldsymbol{\varsigma}_{(s,1)}$ can be retrieved as the preimage $f_\beta^{-1}(c-\mathrm{Sum}(\tilde{\boldsymbol{v}}_{(s,1)}) \pmod{2Ps})$.
\end{IEEEproof}

\begin{example}\label{ex:qVT}
  Assume $\mathcal{J}= \{1,3,4,9,12,13,15,16 \}$, $\boldsymbol{\varsigma}= (4,9,1,12,3,15,16,13)$, $\beta(\boldsymbol{\varsigma})= (3,4,1,5,2,7,8,6)$, and $\mathrm{Sum}(\beta(\boldsymbol{\varsigma})_{(2,1)}) \equiv 6 \pmod{8}$. Suppose $\boldsymbol{\varsigma}$ suffers a burst stable deletion of length $2$ and results in $\boldsymbol{v}= (4,9,1,15,16,13)$. With the knowledge of the set $\mathcal{J}$, we have $\tilde{\boldsymbol{v}}= (3,4,1,7,8,6)$, $\tilde{\boldsymbol{v}}_{(2,1)}= (3,1,8)$, and $\mathrm{Sum}(\tilde{\boldsymbol{v}}_{(2,1)}) \equiv 4 \pmod{8}$. The missing symbol of $\beta(\boldsymbol{\varsigma})_{(2,1)}$ is thus $2$ and the missing symbol of $\boldsymbol{\varsigma}_{(2,1)}$ is $f_\beta^{-1}(2)=3$.
\end{example}

\begin{remark}
Note that Lemma \ref{lem:symbol} also holds when the elements $\{\varsigma_1,\varsigma_2,\ldots,\varsigma_{2Ps}\}$ in $\mathcal{J}$ are not distinct. While the bijection $f_\beta$ does not make sense anymore, there is no problem in finding the missing symbol $t$ in $\beta(\boldsymbol{\varsigma})_{(s,1)}$ and retrieve the missing symbol of $\boldsymbol{\varsigma}_{(s,1)}$ to be the $t$-th smallest element in the (multi-set) $\mathcal{J}$. In fact, let
\begin{equation*}
  \mathcal{J}= \{\underbrace{a_1,\ldots,a_1}_{r_1},\underbrace{a_2,\ldots,a_2}_{r_2},\ldots,\underbrace{a_w,\ldots,a_w}_{r_w}\}
\end{equation*}
with $\sum_{j=1}^w r_j= 2Ps$ and $a_1<a_2<\cdots<a_w$. If $\sum_{j=1}^i r_j +1 \leq t \leq \sum_{j=1}^{i+1} r_j$ for some $i$, then the $t$-th smallest element in the multi-set $\mathcal{J}$ is $a_{i+1}$.
\end{remark}

Lemma \ref{lem:symbol} suggests that, if in Step 2 we find a proper block $\boldsymbol{\varsigma}$ containing the $P+1$ erroneous columns, then the missing symbol of the first row can be retrieved as long as we know the value $c$ such that $\mathrm{Sum}(\beta(\boldsymbol{\varsigma})_{(s,1)})=c \pmod{2Ps}$. The next part describes how to derive such a parameter $c$.

\begin{definition}\label{def:retrieve}
Let $\mathcal{C}_{c_1,c_2}^{\mathrm{retrieve}}(n,s,P)$, with $c_1, c_2 \in \mathbb{Z}_{2Ps}$, be the set of permutations or multi-permutations, such that each codeword $\boldsymbol{\sigma}$ satisfies the following constraints
    \begin{equation}\label{eq:retrieve}
    \begin{cases}
        \begin{aligned}
        \sum\limits_{i=1}^{n/2Ps} \mathrm{Sum}(\beta(\boldsymbol{\sigma}(s,P,2i-1),
        &~ \boldsymbol{\sigma}(s,P,2i))_{(s,1)}) \\
        & \equiv c_1 \pmod{2Ps},
        \end{aligned}\\
        \begin{aligned}
        \sum\limits_{i=1}^{n/2Ps} \mathrm{Sum}(\beta(\boldsymbol{\sigma}(s,P,2i),
        &~ \boldsymbol{\sigma}(s,P,2i+1))_{(s,1)}) \\
        &~~~~~ \equiv c_2 \pmod{2Ps},
        \end{aligned}
    \end{cases}
    \end{equation}
    where $\boldsymbol{\sigma}(s,P,n/Ps+1) = \boldsymbol{\sigma}(s,P,1)$.
\end{definition}

\begin{remark}
In Definition \ref{def:retrieve} we partite $\boldsymbol{\sigma}$ into $\frac{n}{Ps}$ disjoint subarrays, where $\frac{n}{Ps}$ is an even integer.
We combine two subarrays $\boldsymbol{\sigma}(s,P,2i-1),\boldsymbol{\sigma}(s,P,2i)$ together and consider the first row of its permutation projection
$\boldsymbol{v}_{2i-1,2i}\triangleq\beta(\boldsymbol{\sigma}(s,P,2i-1),\boldsymbol{\sigma}(s,P,2i))_{(s,1)}$. Regard $\boldsymbol{v}_{2i-1,2i}$ as a $q$-ary vector of length $2P$ where $q=2Ps$ and we calculate the sum of its entries. In the first equation of Equation (\ref{eq:retrieve}) we do the summation over $\{\boldsymbol{v}_{2i-1,2i}:1\leq i \leq \frac{n}{2Ps}\}$. The second equation is a similar summation over $\{\boldsymbol{v}_{2i,2i+1}:1\leq i \leq \frac{n}{2Ps}\}$.
\end{remark}

\begin{lemma}\label{lem:retrieve}
Suppose $\boldsymbol{\sigma}$ suffers a burst stable deletion of length $s$ and we already find $\boldsymbol{\varsigma}= (\boldsymbol{\sigma}(s,P,k),\boldsymbol{\sigma}(s,P,k+1))$, a block of $2P$ consecutive columns containing all erroneous coordinates. Moreover, if $\boldsymbol{\sigma} \in \mathcal{C}_{c_1,c_2}^{\mathrm{retrieve}}(n,s,P)$, then the value $c$ such that $\mathrm{Sum}(\beta(\boldsymbol{\varsigma})_{(s,1)}) \equiv c \pmod{2Ps}$ can be determined.
\end{lemma}

\begin{IEEEproof}
    To find the value $c$ such that $\mathrm{Sum}(\beta(\boldsymbol{\varsigma})_{(s,1)}) \equiv c \pmod{2Ps}$, we turn to either the first or the second equation in Equation (\ref{eq:retrieve}), depending on the parity of $k$. WLOG let $k$ be odd, then by the first equation, we have
    \begin{align*}
            c
            & = \mathrm{Sum}(\beta(\boldsymbol{\varsigma})_{(s,1)}) \\
            & \equiv c_1- \sum\limits_{i \neq (k+1)/2 } \mathrm{Sum}(\boldsymbol{v}_{2i-1,2i}) \pmod{2Ps},
    \end{align*}
    where $\boldsymbol{v}_{2i-1,2i}\triangleq\beta(\boldsymbol{\sigma}(s,P,2i-1),\boldsymbol{\sigma}(s,P,2i))_{(s,1)}$, and thus the value $c$ is determined.
\end{IEEEproof}

\begin{example}\label{ex:retrieve}
    Consider the set $\mathcal{C}_{6,2}^{\mathrm{retrieve}}(16,2,2)$, suppose $\boldsymbol{\sigma}=(7,8,2,5,4,9,1,12,3,15,16,13,14,6,11,10)$ (from the set) suffers a burst stable deletion of length $2$ and the resultant vector is $\boldsymbol{\tau}=(7,8,2,5,4,9,1,15,16,13,14,6,11,10)$. In addition, suppose that we have already known that the erroneous coordinate in the first row belongs to the interval $[4,5]$. Then we can find the block $\boldsymbol{\varsigma}= (\boldsymbol{\sigma}(2,2,2),\boldsymbol{\sigma}(2,2,3))$ containing the three possible erroneous columns, and calculate the parameter $c$ such that $\mathrm{Sum}(\beta(\boldsymbol{\varsigma})_{(2,1)}) \equiv c \pmod{8}$, by $c= 2- \mathrm{Sum}(\beta(\boldsymbol{\sigma}(2,2,4),\boldsymbol{\sigma}(2,2,1))_{(2,1)}) \equiv 6 \pmod{8}$.
\end{example}

\subsection{Summary of this section}

In this section, we introduce a different strategy to retrieve the missing symbol of $\boldsymbol{\sigma}_{(s,1)}$.
We summarize this section into the following theorem, which will be the key building block for our codes in the next two sections.

\begin{theorem} \label{thm:symbol}
Consider a permutation (multi-permutation) $\boldsymbol{\sigma}\in \mathcal{C}_{c_1,c_2}^{\mathrm{retrieve}}(n,s,P)$ and assume that $\boldsymbol{\sigma}$ suffers a burst stable deletion of length $s$. If the erroneous coordinate of $\boldsymbol{\sigma}_{(s,1)}$ is within a known interval of length $P$, then the missing symbol of $\boldsymbol{\sigma}_{(s,1)}$ can be exactly retrieved.
\end{theorem}

\begin{IEEEproof}
Since the erroneous coordinate of $\boldsymbol{\sigma}_{(s,1)}$ is within a known interval of length $P$, the erroneous coordinates on the other rows are within some $P+1$ columns. By Step 2, we can find a proper block $\boldsymbol{\varsigma}= (\boldsymbol{\sigma}(s,P,k),\boldsymbol{\sigma}(s,P,k+1))$ of $2P$ consecutive columns containing all erroneous coordinates. By Lemma \ref{lem:retrieve} in Step 3, the value $c$ such that $\mathrm{Sum}(\beta(\boldsymbol{\varsigma})_{(s,1)}) \equiv c \pmod{2Ps}$ can be determined since $\boldsymbol{\sigma} \in \mathcal{C}_{c_1,c_2}^{\mathrm{retrieve}}(n,s,P)$. Then by Lemma \ref{lem:symbol} in Step 3, the missing symbol of $\boldsymbol{\varsigma}_{(s,1)}$ (and thus $\boldsymbol{\sigma}_{(s,1)}$) can be retrieved.
\end{IEEEproof}

\section{$s$-BSD permutation/multi-permutation codes}\label{Sec:fixed}

In this section we consider the fixed length burst model and present our constructions of $s$-BSD permutation and multi-permutation codes. Our code has $\log {n}+2\log  \log  n +O(1)$ bits of redundancy, which outperforms the previously best known result from \cite{Chee-20-IT-PBSD} with $2\log  n+O(1)$ bits of redundancy. The following definition is needed.

\begin{definition}\label{def:permutation rank}
For two distinct permutations $\boldsymbol{\sigma}, \boldsymbol{\pi} \in S_n$, define the partial order $\boldsymbol\sigma \preceq \boldsymbol\pi$ if there exists some $1\le j \le n$ such that $\sigma_i=\pi_i$ for $1\le i < j$ and $\sigma_j < \pi_j$. This is a total order in $S_n$ known as the \emph{lexicographic order}. Let $\ell(\boldsymbol{\sigma})$ be the \emph{lexicographic rank} of $\boldsymbol{\sigma} \in S_n$, i.e., $\boldsymbol{\sigma}$ is the $\ell(\boldsymbol{\sigma})$-th smallest permutation in the poset $(S_n,\preceq)$.
For a vector $\boldsymbol{u}$ of length $n$ with positive integer entries, recall that $\beta(\boldsymbol{u})$ is the permutation projection of $\boldsymbol{u}$ and we define the \emph{permutation rank} of $\boldsymbol{u}$ as $\mu(\boldsymbol{u}) = \ell(\beta(\boldsymbol{u}))$.
\end{definition}

\begin{remark}\label{rmk:recovery}
Note that we can recover an integer vector $\boldsymbol{u}$ given the knowledge of its permutation rank $\mu(\boldsymbol{u})$ and the multiset of all its entries $\{u_i : 1 \leq i \leq n\}$. For example, if a vector $\boldsymbol{u}=(u_1,u_2,u_3,u_4)$ has permutation rank $\mu(\boldsymbol{u}) = 2$, then $\beta(\boldsymbol{u}) = (1,2,4,3)$. Furthermore, if $\{u_1,u_2,u_3,u_4\} = \{1,3,5,6\}$, then $\boldsymbol{u} = (1,3,6,5)$.
\end{remark}

We now present our codes against fixed length burst stable deletions. Subsection \ref{subsec:fixedperm} deals with permutations and Subsection \ref{subsec:fixedmulti} deals with multi-permutations.

\subsection{$s$-BSD permutation codes} \label{subsec:fixedperm}

\begin{definition}\label{def:good}
Consider a permutation $\boldsymbol{\sigma} \in S_n$ written in an array of size $s\times \frac{n}{s}$. Let $\alpha(\boldsymbol{\sigma}_{(s,1)})$ be the signature of its first row. If $\alpha(\boldsymbol{\sigma}_{(s,1)})$ is a binary vector with maximal runlength at most $P-1$, then we say $\boldsymbol{\sigma}$ is a \emph{good permutation}.
\end{definition}

\begin{lemma}\label{lem:RLLperm}
The number of good permutations is at least $\frac{n!}{2}$.
\end{lemma}

\begin{IEEEproof}
We prove the lemma by a probabilistic analysis. Uniformly choose a random permutation $\boldsymbol{\sigma}$ in $S_n$. Then, for each $i \in [1,\frac{n}{s}-P]$, let $\mathbb{E}_i$ be the event that $\left( \alpha(\boldsymbol{\sigma}_{(s,1)})_{i},\alpha(\boldsymbol{\sigma}_{(s,1)})_{i+1}, \ldots, \alpha(\boldsymbol{\sigma}_{(s,1)})_{i+P-1} \right)$ is either $\boldsymbol{0}^P$ or $\boldsymbol{1}^P$, i.e., either $\boldsymbol{\sigma}_{(i-1)s+1}> \boldsymbol{\sigma}_{is+1}> \cdots > \boldsymbol{\sigma}_{(i+P-1)s+1}$ or $\boldsymbol{\sigma}_{(i-1)s+1}< \boldsymbol{\sigma}_{is+1}< \cdots < \boldsymbol{\sigma}_{(i+P-1)s+1}$. The probability of $\mathbb{E}_i$ can be calculated by

\begin{equation*}
\mathbf{Pr}(\mathbb{E}_i) = \frac{2} {(P+1)!} \leq \frac{2}{2^P} \leq \frac{s}{2n}.
\end{equation*}

By the union bound, the probability of the event that $\boldsymbol{\sigma}$ is not a good permutation is upper bounded by
\begin{align*}
&~~~~ \mathbf{Pr} \left( \boldsymbol{\sigma} \text{ is not a good permutation} \right) \\
& \leq \sum_{i=1}^{\frac{n}{s}-P} \mathbf{Pr}(\mathbbm{E}_i)  \leq \frac{s}{2n} \left( \frac{n}{s}-P \right) \leq \frac{1}{2},
\end{align*}
and thus the number of good permutations is at least $\frac{n!}{2}$.
\end{IEEEproof}

Following Lemma \ref{lem:RLLperm}, we consider good permutations $\boldsymbol{\sigma}$ with $\alpha(\boldsymbol{\sigma}_{(s,1)})$ following a binary VT constraint. When a burst stable deletion of length $s$ occurs in  $\boldsymbol{\sigma}$, the first row suffers exactly one stable deletion and so does its signature. By decoding the first row of its signature, the erroneous coordinate of $\alpha(\boldsymbol{\sigma}_{(s,1)})$ is bounded in an interval of length $P-1$. Now we explain the relation of the erroneous coordinate between a sequence and its signature.

\begin{lemma}[Lemma 2, \cite{Saeki-18-ISIT-BD}]\label{lem:relation}
For a vector $\boldsymbol{x}$ of length $n$, either $\alpha(\boldsymbol{x}_{[n] \setminus \{i\}})= \alpha(\boldsymbol{x})_{[n-1] \setminus \{i-1\}}$
or $\alpha(\boldsymbol{x}_{[n] \setminus \{i\}})= \alpha(\boldsymbol{x})_{[n-1] \setminus \{i\}}$ holds. That is, when the erroneous coordinate of a sequence is $i$, the erroneous coordinate of its signature is either $i-1$ or $i$.
\end{lemma}

By Lemma \ref{lem:relation} we have the following lemma.
\begin{lemma}\label{lem:P-bound}
If a good permutation $\boldsymbol{\sigma}$ suffers a burst deletion of length $s$ and $\alpha(\boldsymbol{\sigma}_{(s,1)})$ follows a binary VT constraint, the erroneous coordinate of $\boldsymbol{\sigma}_{(s,1)}$ can be located in an interval of length $P$.
\end{lemma}

Up till now, we have used good permutations to finish Step 1 mentioned in the previous section. Then we can proceed with Steps 2 and 3 to retrieve the missing symbol of $\boldsymbol{\sigma}_{(s,1)}$. Afterwards, some other constraint on the first row can help us exactly locate the  erroneous coordinate of $\boldsymbol{\sigma}_{(s,1)}$. Then the erroneous coordinates on the other rows are within two adjacent columns. The correction of these two columns follows the same way as in \cite{Chee-20-IT-PBSD} and is rephrased here for completeness.

\begin{definition}\label{def:encode_columns}
 Let $\mathcal{C}_{d_1,d_2}^{\mathrm{recover}}(n,s)$, with $d_1,d_2 \in \mathbb{Z}_{(2s)!}$, be the set of all permutations $\boldsymbol{\sigma} \in S_n$ which satisfies the following constraints
 \begin{equation}\label{eq:recover}
        \begin{cases}            \sum\limits_{i=1}^{n/2s} \mu ( \boldsymbol{\sigma}^{(s,2i-1)}, \boldsymbol{\sigma}^{(s,2i)}) = d_1 \pmod{(2s)!} \\
        \sum\limits_{i=1}^{n/2s} \mu ( \boldsymbol{\sigma}^{(s,2i)}, \boldsymbol{\sigma}^{(s,2i+1)} ) = d_2 \pmod{(2s)!}
        \end{cases},
    \end{equation}
where $\boldsymbol{\sigma}^{(s,n/2s+1)} = \boldsymbol{\sigma}^{(s,1)}$.
\end{definition}

\begin{lemma}\label{lem:recover}
Suppose we have located all erroneous coordinates in two adjacent columns. If $\boldsymbol{\sigma} \in \mathcal{C}_{d_1,d_2}^{\mathrm{recover}}(n,s)$, then the two erroneous columns can be corrected by its permutation rank derived from Equation (\ref{eq:recover}) and the set of its entries.
\end{lemma}

\begin{example}\label{ex:recover}
    Consider the set $\mathcal{C}_{2,3}^{\mathrm{recover}}(16,2)$, suppose $\boldsymbol{\sigma}=(7,8,2,5,4,9,1,12,3,15,16,13,14,6,11,10)$ (from the set) suffers a burst stable deletion of length $2$. If the resultant vector is $\boldsymbol{\tau}=(7,8,2,5,4,9,1,15,16,13,14,6,11,10)$ and the erroneous coordinate of the first row is $5$. All we need is to recover the columns indexed by $4$ and $5$ and it can be done as follows. Firstly, we know that $\boldsymbol{\sigma}^{(2,1)}= (7,8)$, $\boldsymbol{\sigma}^{(2,2)}= (2,5)$, $\boldsymbol{\sigma}^{(2,3)}= (4,9)$, $\boldsymbol{\sigma}^{(2,6)}= (16,13)$, $\boldsymbol{\sigma}^{(2,7)}= (14,6)$, ,$\boldsymbol{\sigma}^{(2,8)}= (11,10)$, the four elements in $\boldsymbol{\sigma}^{(2,4)}$ and $\boldsymbol{\sigma}^{(2,5)}$ are $\{1,3,12,15\}$. Then by Equation (\ref{eq:recover}) we obtain $\mu(\boldsymbol{\sigma}^{(2,4)},\boldsymbol{\sigma}^{(2,5)})= 3- \sum_{j \in \{1,3,4\}} \mu(\boldsymbol{\sigma}^{(2,2j)},\boldsymbol{\sigma}^{(2,2j+1)})= 3 \pmod{24}$. Then by Remark \ref{rmk:recovery}, the permutation projection is $\beta(\boldsymbol{\sigma}^{(2,4)},\boldsymbol{\sigma}^{(2,5)})=(1,3,2,4)$ and thus $(\boldsymbol{\sigma}^{(2,4)},\boldsymbol{\sigma}^{(2,5)})= (1,12,3,15)$.
\end{example}

Summing up the above, the full construction of our $s$-BSD permutation codes is as follows.

\begin{construction}\label{constr:perm_burst}
Let $a \in \mathbb{Z}_{n/s}$, $c_1,c_2 \in \mathbb{Z}_{2Ps}$, and $d_1,d_2 \in \mathbb{Z}_{(2s)!}$.
Define the code $\mathcal{C}_s^1 \triangleq \mathcal{C}_s^1(n;a;c_1,c_2;d_1,d_2)$ as
\begin{align*}
    \mathcal{C}_s^1= \big\{ \boldsymbol{\sigma} \in S_n:
    &~\alpha(\boldsymbol{\sigma}_{(s,1)}) \in VT_a(n/s-1), \boldsymbol{\sigma} \text{ is good}, \\
    &~~~~\boldsymbol{\sigma} \in \mathcal{C}_{c_1,c_2}^{\mathrm{retrieve}}(n,s,P) \cap \mathcal{C}_{d_1,d_2}^{\mathrm{recover}}(n,s) \big\}.
\end{align*}
\end{construction}

\begin{theorem}\label{thm:BSD}
The permutation code $\mathcal{C}_s^1(n;a;c_1,c_2;d_1,d_2)$ is an $s$-BSD permutation code over $S_n$.
Moreover, by choosing proper parameters there is an $s$-BSD permutation code with redundancy at most
$\log  n+ 2\log \log  n+O(1)$.
\end{theorem}

\begin{IEEEproof}
Suppose $\boldsymbol{\sigma} = \left( \sigma_1,\sigma_2,\ldots,\sigma_n \right) \in \mathcal{C}_s^1$ suffers a burst stable deletion of length $s$. Since $\alpha(\boldsymbol{\sigma}_{(s,1)}) \in VT_a(n/s-1)$ and $\boldsymbol{\sigma}$ is a good permutation, by Lemma \ref{lem:P-bound} the erroneous coordinate of $\boldsymbol{\sigma}_{(s,1)}$ can be located in an interval of length at most $P$.
Moreover, since $\boldsymbol{\sigma} \in \mathcal{C}_{c_1,c_2}^{\mathrm{retrieve}}(n,s,P)$, by Theorem \ref{thm:symbol} we can retrieve the missing symbol of $\boldsymbol{\sigma}_{(s,1)}$.
Note that the erroneous coordinate of $\alpha(\boldsymbol{\sigma}_{(s,1)})$ lies in a consecutive run of 0's or 1's, which means that the erroneous coordinate of $\boldsymbol{\sigma}_{(s,1)}$ lies in a monotone decreasing sequence or a monotone increasing sequence. In such a sequence, the coordinate of the known missing symbol is certainly unique. Therefore, the erroneous coordinate of $\boldsymbol{\sigma}_{(s,1)}$ can be exactly located. Finally, the two erroneous columns can be corrected by Lemma \ref{lem:recover}.

By the pigeon-hole principal, there exist parameters $a \in \mathbb{Z}_{n/s}$, $c_1,c_2 \in \mathbb{Z}_{2Ps}$, $d_1,d_2 \in \mathbb{Z}_{(2s)!}$ such that the redundancy of $\mathcal{C}_s^1$ is at most
\begin{align*}
&~~~\log  2 + \log  {\frac{n}{s}}+ 2 \log  (2Ps)+ 2 \log  (2s)!\\
&= \log  n+ 2\log \log  n+O(1),
\end{align*}
where the first term is the 1 bit redundancy due to the size of good permutations (Lemma \ref{lem:RLLperm}).
\end{IEEEproof}

\begin{example}\label{ex:burst_deletion}
    Consider the set $\mathcal{C}_2^1(16;3;6,2;2,3)$, suppose
    $\boldsymbol{\sigma}=(7,8,2,5,4,9,1,12,3,15,16,13,14,6,11,10)$ (from the set) suffers a burst stable deletion of length $2$ and the resultant vector is $\boldsymbol{\tau}=(7,8,2,5,4,9,1,15,16,13,14,6,11,10)$. Firstly, since $\alpha(\boldsymbol{\sigma}_{(2,1)}) \in VT_3(7)$, we can use a VT decoder to recover $\alpha(\boldsymbol{\sigma}_{(2,1)})= (0,1,0,1,1,0,0)$ from $\alpha(\boldsymbol{\tau}_{(2,1)})= (0,1,0,1,0,0)$, and locate the erroneous coordinate in the first row in the interval $[4,5]$. Then we find a proper block $\boldsymbol{\varsigma}= (\boldsymbol{\sigma}(2,2,2),\boldsymbol{\sigma}(2,2,3))$ and by Example \ref{ex:retrieve},  we obtain $\mathrm{Sum}(\beta(\boldsymbol{\varsigma})_{(2,1)}) \equiv 6 \pmod{8}$. By Example \ref{ex:qVT}, we retrieve the missing symbol of $\boldsymbol{\sigma}_{(s,1)}$ to be $3$. Now we use an $n$-ary VT decoder to recover $\boldsymbol{\sigma}_{(2,1)}$ and locate the exact erroneous coordinate. Finally, all erroneous coordinates are located in two adjacent columns and we recover these two columns by Example \ref{ex:recover}, which completes the decoding process.
\end{example}

\subsection{$s$-BSD multi-permutation codes}\label{subsec:fixedmulti}

Now we consider $s$-BSD multi-permutation codes in $S_n^{\boldsymbol{r}}$, where $\boldsymbol{r}=(r_1,r_2,\dots,r_w)$ is the given multiplicity vector.
Similarly as the permutation codes, we can recover the first row of a multi-permutation $\boldsymbol{\sigma}\in S_n^{\boldsymbol{r}}$.
However note that in a multi-permutation there are repeated entries, and thus recovering $\boldsymbol\sigma_{(s,1)}$ can only help us locate the erroneous coordinate within an interval of length $r$ in the worst case, where $r= \max_{1 \leq i \leq w} r_i$. Thus, all the erroneous coordinates in the other rows are within at most $r+1$ columns. To deal with this issue we can reconsider $\boldsymbol{\sigma}$ in an array of size $s(r+1) \times \frac{n}{s(r+1)}$ and then all the erroneous coordinates are within at most two adjacent columns in this new array, and the correction of these two columns can be done in a similar way as Lemma \ref{lem:recover}.

In the constructions below we start with the case where $n=wr$ and $r_i=r$ for all $1\leq i \leq w$ (i.e., $\boldsymbol{r}$ is regular), and then for the irregular case. Moreover, $w$ is assumed to be even for simplicity. Recall that we have fixed the parameter $P=\lceil \log {\frac{4n}{s}} \rceil$. First we view a multi-permutation as an array of size $s\times \frac{n}{s}$. We say that a multi-permutation $\boldsymbol{\sigma}$ is \emph{good} if $\alpha(\boldsymbol{\sigma}_{(s,1)})$ is a binary vector with maximal runlength at most $P-1$. Similarly, by a probabilistic analysis, the number of good multi-permutations is at least half the size of $S_n^{\boldsymbol{r}}$.

\begin{lemma}\label{lem:RLLmltiperm}
The number of good multi-permutations is at least $\frac{n!}{2(r!)^w}$ when $P+1 \geq 2^{2r}$.
\end{lemma}

\begin{IEEEproof}
Uniformly choose a random multi-permutation $\boldsymbol{\sigma}\in S_n^{\boldsymbol{r}}$. For each $i \in [1,\frac{n}{s}-P]$, let $\mathbb{E}_i$ be the event that $\left( \alpha(\boldsymbol{\sigma}_{(s,1)})_{i},\alpha(\boldsymbol{\sigma}_{(s,1)})_{i+1}, \ldots, \alpha(\boldsymbol{\sigma}_{(s,1)})_{i+P-1} \right)$ is either $\boldsymbol{0}^{P}$ or $\boldsymbol{1}^{P}$, i.e., either $\boldsymbol{\sigma}_{(i-1)s+1} \geq \boldsymbol{\sigma}_{is+1} \geq \cdots \geq \boldsymbol{\sigma}_{(i+P-1)s+1}$ or $\boldsymbol{\sigma}_{(i-1)s+1} \leq \boldsymbol{\sigma}_{is+1} \leq \cdots \leq \boldsymbol{\sigma}_{(i+P-1)s+1}$. For simplicity, we assume that $(P+1)$ is a multiple of $r$, and the probability of $\mathbb{E}_i$ can be calculated by

\begin{align*}
    \mathbf{Pr}(\mathbb{E}_i)
    &= \frac{\sum\limits_{\substack{t_1+\cdots+t_w= P+1 \\ 0 \leq t_1, \ldots, t_w \leq r}} 2} {\sum\limits_{\substack{t_1+\cdots+t_w= P+1 \\ 0 \leq t_1, \ldots, t_w \leq r}} \frac{(P+1)!}{\prod\limits_{1 \leq j \leq w} t_j!}} \\
    & \overset{(\star)}{\leq} \frac{2}{2^{P-2r}(P+1)} \\
    & \overset{(\ast)}{\leq} \frac{1}{2^{P-1}}  \leq \frac{s}{2n}.
\end{align*}
Note that for any $t_1, t_2 \in [0,r]$, we have
\begin{equation*}
  (t_1)!(t_2)! \leq
  \begin{cases}
    (t_1+t_2)!, & \mbox{if } t_1+t_2 \leq r;\\
    r!(t_1+t_2-r)!, & \mbox{otherwise}.
  \end{cases}
\end{equation*}
Thus, ($\star$) holds since $\prod\limits_{1 \leq j \leq w} t_j! \leq (r!)^{\frac{P+1}{r}} \leq (2r)!r^{P-2r}$ for any choice of integers $t_1, \ldots , t_w$ such that $0 \leq t_1, \ldots , t_w \leq r$ and $t_1+\cdots+t_w= P+1$, and $\frac{(P+1)!}{(2r)!r^{P-2r}} \geq 2^{P-2r}(P+1)$. ($\ast$) holds when $P+1 \geq 2^{2r}$.

By the union bound, the probability of the event that $\boldsymbol{\sigma}$ is not a good multi-permutation is upper bounded by
\begin{align*}
&~~~~\mathbf{Pr} \left( \boldsymbol{\sigma} \text{ is not a good multi-permutation} \right) \\
& \leq \sum_{i=1}^{\frac{n}{s}-P} \mathbf{Pr}(\mathbbm{E}_i) \leq \frac{s}{2n} \left( \frac{n}{s}-P \right) \leq \frac{1}{2},
\end{align*}
and thus the number of good multi-permutations is at least $\frac{n!}{2(r!)^w}$.
\end{IEEEproof}

The full construction of our $s$-BSD multi-permutation codes is as follows. Note that the code $\mathcal{C}_{d_1,d_2}^{\mathrm{recover}}(n,s)$ in Definition \ref{def:encode_columns} can be naturally generalized to multi-permutations.

\begin{construction}\label{constr:multiperm_burst}
Let $n=rw=s(r+1)t$ where $t$ is even.
Let $a \in \mathbb{Z}_{n/s}$, $c_1,c_2 \in \mathbb{Z}_{2Ps}$, and $d_1,d_2 \in \mathbb{Z}_{(2sr)!}$.
Define the code $\mathcal{C}_s^2 \triangleq \mathcal{C}_s^2(n;a;c_1,c_2;d_1,d_2)$ as
\begin{align*}
    \mathcal{C}_s^2= \big\{ \boldsymbol{\sigma} \in
    &~ S_n^{\boldsymbol{r}}:\alpha(\boldsymbol{\sigma}_{(s,1)}) \in VT_a(n/s-1), \boldsymbol{\sigma} \text{ is good}, \\
     & \boldsymbol{\sigma} \in \mathcal{C}_{c_1,c_2}^{\mathrm{retrieve}}(n,s,P) \cap \mathcal{C}_{d_1,d_2}^{\mathrm{recover}}(n,s(r+1)) \big\}.
\end{align*}
\end{construction}

\begin{theorem}\label{thm:multiprem_burst}
The code $\mathcal{C}_s^2$ is an $s$-BSD multi-permutation code over $S_n^{\boldsymbol{r}}$.
Moreover, by choosing proper parameters there is an $s$-BSD multi-permutation code over $S_n^{\boldsymbol{r}}$ with redundancy at most $\log  n +2\log  \log  n +O(1)$.
\end{theorem}

\begin{IEEEproof}
Suppose $\boldsymbol{\sigma} = \left( \sigma_1,\sigma_2,\ldots,\sigma_n \right) \in \mathcal{C}_s^2$ suffers a burst stable deletion of length $s$.
Since $\alpha(\boldsymbol{\sigma}_{(s,1)}) \in VT_a(n/s-1)$ and $\boldsymbol{\sigma}$ is a good multi-permutation, by Lemma \ref{lem:P-bound} the erroneous coordinate of $\boldsymbol{\sigma}_{(s,1)}$ can be located in an interval of length at most $P$.
Moreover, since $\boldsymbol{\sigma} \in \mathcal{C}_{c_1,c_2}^{\mathrm{retrieve}}(n,s,P)$, by Theorem \ref{thm:symbol} we can retrieve the missing symbol of $\boldsymbol{\sigma}_{(s,1)}$. By an $n$-ary VT decoder, we can recover $\boldsymbol{\sigma}_{(s,1)}$.
However, due to the existence of repeated entries in $\boldsymbol{\sigma}_{(s,1)}$, recovering $\boldsymbol{\sigma}_{(s,1)}$ only helps locate the erroneous coordinate within an interval of length $r$, in the worst case. Finally, by viewing the multi-permutation in an array of size $s(r+1)\times t$, all the erroneous coordinates are within two adjacent columns. Since $\boldsymbol{\sigma} \in \mathcal{C}_{d_1,d_2}^{\mathrm{recover}}(n,s(r+1))$, the correction can be done in a similar way as Lemma \ref{lem:recover}.

The code redundancy can be calculated by the pigeon-hole principle immediately.
\end{IEEEproof}

To construct codes for irregular multi-permutations, simply choose $r_{\mathrm{max}} = \max_{i \in [w]} r_i$ and apply the above construction.

\begin{corollary}
There exists an $s$-BSD multi-permutation code in $S_n^{\boldsymbol{r}}$ (here $\boldsymbol{r}$ could be irregular) with redundancy at most $\log  n +2\log  \log  n +O(1)$.
\end{corollary}

\section{$^{\leq}s$-BSD permutation/multi-permutation correcting codes}\label{Sec:variable}

Now we move on to the variable length burst. A trivial approach is to take the intersection of our $s'$-BSD permutation/multi-permutation codes for all $1 \leq s' \leq s$ and the redundancy is at most $s \log  n +O(\log  \log  n)$, which is better than the best known result $(2s-1)\log  n+O(1)$ from \cite{Chee-20-IT-PBSD}.
In this section, we show that we can further reduce the redundancy to $\log  {n}+O(\log  \log  n)$.  First we recall some terminologies in \cite{Lenz-20-ISIT-BD} used for binary $^{\leq}s$-burst deletion correcting codes.

\begin{definition}
For a binary vector $\boldsymbol{x}$ of length $n$, its
$\boldsymbol{p}$-\emph{indicator vector} $\mathbbm{1}_{\boldsymbol{p}}(\boldsymbol{x})$, where $\boldsymbol{p} \in \Sigma_2^m$ with $m \leq n$, is a vector of length $n$ defined as
\begin{equation*}
    \mathbbm{1}_{\boldsymbol{p}}(\boldsymbol{x})_i =
        \begin{cases}
            1, & \text{if } \boldsymbol{x}_{[i,i+m-1]} = \boldsymbol{p}, 1\leq  i \leq n-m+1; \\
            0, & \text{otherwise}.
        \end{cases}
\end{equation*}
Furthermore, let $n_{\boldsymbol{p}}(\boldsymbol{x})$ be the number of ones in $\mathbbm{1}_{\boldsymbol{p}}(\boldsymbol{x})$ and $\alpha_{\boldsymbol{p}}(\boldsymbol{x})$ be a vector of length $n_{\boldsymbol{p}}(\boldsymbol{x}) + 1$ whose $i$-th entry is the distance between coordinates of the $i$-th and $(i+1)$-th $1$ in the vector $(1, \mathbbm{1}_{\boldsymbol{p}}(\boldsymbol{x}),1)$.
\end{definition}

\begin{definition}
Let $\boldsymbol{p} \in \Sigma_2^m$ and $\delta > m$ be a positive integer. A vector $\boldsymbol{x}$ is called a $(\boldsymbol{p},\delta)$-\emph{dense vector}, if each interval of length $\delta$ in $\boldsymbol{x}$ contains at least one consecutive subsequence $\boldsymbol{p}$, i.e., for each $i \in [n - \delta + 1]$, there exists $j \in [i,i + \delta - m]$, such that $\boldsymbol{p} = \left( x_j,x_{j+1},\ldots,x_{j+m-1} \right)$. Let $\mathcal{D}_{\boldsymbol{p},\delta}^{n}$ denote all $(\boldsymbol{p},\delta)$-dense vectors in $\Sigma_2^n$.
\end{definition}

In the rest of the paper we set $\boldsymbol{p} = 0^s 1^s$ and fix $\delta = s 2^{2s+1} \lceil \log {n} \rceil$.
Note that the structure of $\boldsymbol{p}$ indicates that $\alpha_{\boldsymbol{p}}(\boldsymbol{x})_i \geq 2s$ for $i \geq 2$. Furthermore, $\alpha_{\boldsymbol{p}}(\boldsymbol{x})_i \leq \delta$ for any $i \in [n_{\boldsymbol{p}}(\boldsymbol{x}) + 1]$ if and only if $\boldsymbol{x}$ is a $(\boldsymbol{p},\delta)$-dense vector.

\begin{lemma}[Lemma 2, \cite{Lenz-20-ISIT-BD}]\label{lem:dense}
For any integer $a_1 \in \mathbb{Z}_4$ and $a_2 \in \mathbb{Z}_{2n}$, let $\mathcal{C}_{a_1,a_2}^{\mathrm{locate}}(n,s)$ be the set
\begin{align*}
    \big\{\boldsymbol{x} \in \Sigma_2^n : \boldsymbol{x} \in \mathcal{D}_{\boldsymbol{p},\delta}^{n},
    &~ n_{\boldsymbol{p}}(\boldsymbol{x}) = a_1 \pmod{4}, \\
    & \mathrm{VT}(\alpha_{\boldsymbol{p}}(\boldsymbol{x})) = a_2 \pmod{2n} \big\}.
\end{align*}
 Suppose $\boldsymbol{x} \in \mathcal{C}_{a_1,a_2}^{\mathrm{locate}}(n,s)$ suffers a burst deletion of length $s'$ where $1 \leq s' \leq s$, the decoder can find an interval of length at most $\delta+s'-1$ which contains all erroneous coordinates of $\boldsymbol{x}$.
\end{lemma}

In order to borrow Lemma \ref{lem:dense} to locate all erroneous coordinates of a permutation or a multi-permutation in a bounded interval, we need a map converting permutations and multi-permutations to binary vectors which satisfies the following constraints.
\begin{itemize}
  \item[I.] When a permutation (multi-permutation) suffers a burst stable deletion of length $s$ ($s \geq 2$), the corresponding binary vector also suffers a burst deletion of length $s$.
  \item[II.] The permutations or multi-permutations whose corresponding binary vector is $(\boldsymbol{p},\delta)$-dense account for a large proportion of the set $S_n$ or $S_n^{\boldsymbol{r}}$ (and they are referred to as $(\boldsymbol{p},\delta)$-\emph{dense permutations/multi-permutations}).
\end{itemize}

The parity indicator vector defined below is a map satisfying these two constraints.

\begin{definition}
For an integer vector $\boldsymbol{\sigma}$, its \emph{parity indicator vector} $\mathbbm{1}(\boldsymbol{\sigma})$ is defined by
\begin{equation*}
\mathbbm{1}(\boldsymbol{\sigma})_i =
\begin{cases}
1, & \text{if } \sigma_i \text{ is odd}; \\
0, & \text{if } \sigma_i \text{ is even}.
\end{cases}
\end{equation*}
\end{definition}

It is routine to check that when $\boldsymbol{\sigma}$ suffers a burst stable deletion of length $s$ ($s \geq 2$), its parity indicator vector also suffers a burst deletion of length $s$. To verify Constraint II we prove the following lemma for permutations. The generalization to multi-permutations is straightforward.

\begin{lemma}\label{lem:densesize}
The number of $(\boldsymbol{p},\delta)$-dense permutations is at least $\frac{n!}{2}$.
\end{lemma}

\begin{IEEEproof}
Uniformly choose a random permutation $\boldsymbol{\sigma}\in S_n$. Let $\mathbb{E}_i$ be the event that $(\mathbbm{1}(\boldsymbol{\sigma})_{i},\mathbbm{1}(\boldsymbol{\sigma})_{i+1},\ldots,\mathbbm{1}(\boldsymbol{\sigma})_{i+\delta-1})$ does not contain $\boldsymbol{p}=0^s1^s$ as a consecutive subsequence, $i \in [1,n-\delta+1]$. The probability of $\mathbb{E}_i$ can be calculated by

\begin{equation*}
\begin{aligned}
\mathbf{Pr} (\mathbb{E}_i)
        &\leq \prod_{j=0}^{\frac{\delta}{2s}-1} \mathbf{Pr} (( \mathbbm{1}(\boldsymbol{\sigma})_{[i+2sj,i+2s(j+1)-1]}) \neq \boldsymbol{p})\\
        & = \left(1- \frac{\binom{n/2}{s}^2 \cdot (s!)^2}{\binom{n}{2s} \cdot (2s)!} \right)^{\frac{\delta}{2s}} \\
        & = \left( 1 - \frac{1}{2^{2s}} \cdot  \frac{n(n-2)\cdots (n-2s+2)}{(n-1)(n-3) \cdots (n-2s+1)} \right)^{\frac{\delta}{2s}} \\
        & \leq \left( 1 - \frac{1}{2^{2s}} \right)^{\frac{\delta}{2s}} \\
        & = \exp \left\{ {\frac{\delta}{2s}} \ln{ \left( 1 - \frac{1}{2^{2s}} \right)} \right\} \\
        & \overset{(\star)}{\leq} \exp \left\{  - \frac{1}{2^{2s}} 2^{2s} \lceil \log {n} \rceil \right\} \leq \frac{1}{n^{\log e}} \leq \frac{1}{2n},
\end{aligned}
\end{equation*}
where ($\star$) holds since $\ln{(1+x)} \leq x$ for all $x \in \mathbb{R}$.

By the union bound, the probability of the event that $\boldsymbol{\sigma}$ is not a $(\boldsymbol{p},\delta)$-dense permutation is upper bounded by
\begin{align*}
&~~~~ \mathbf{Pr} \left( \boldsymbol{\sigma} \text{ is not a } (\boldsymbol{p},\delta) \text{-dense permutation} \right) \\
& \leq \sum_{i=1}^{n-\delta+1} \mathbf{Pr}(\mathbbm{E}_i) \leq n \cdot \frac{1}{2n} = \frac{1}{2},
\end{align*}
and thus the number of $(\boldsymbol{p},\delta)$-dense permutations is at least $\frac{n!}{2}$.
\end{IEEEproof}

We are now ready to present our codes against variable length burst stable deletions. Subsection \ref{subsec:varperm} deals with permutations and Subsection \ref{subsec:varmulti} deals with multi-permutations.

\subsection{$^\leq s$-BSD permutation codes}\label{subsec:varperm}

Given $s$ and $P$, we define the PSVT codes as follows (which means the SVT version for permutations).

\begin{construction}\label{con:PSVT}
Let $b_1 \in \mathbb{Z}_{P}$, $b_2 \in \mathbb{Z}_2$, and $c_1,c_2 \in \mathbb{Z}_{2Ps}$, define the code $\mathrm{PSVT}_{b_1,b_2,c_1,c_2}(n,s,P)$ as
\begin{align*}
   \big\{ \boldsymbol{\sigma} \in S_n:
   &~ \boldsymbol{\sigma} \in \mathcal{C}_{c_1,c_2}^{\mathrm{retrieve}}(n,s,P), \\
   & \alpha(\boldsymbol{\sigma}_{(s,1)}) \in \mathrm{SVT}_{b_1,b_2}(n,P) \big\}.
\end{align*}
\end{construction}

\begin{lemma}\label{lem:PSVT}
Suppose an arbitrary permutation $\boldsymbol\sigma$ from $\mathrm{PSVT}_{b_1,b_2,c_1,c_2}(n,s,P)$ suffers a burst stable deletion of length $s$ and the erroneous coordinate of $\boldsymbol{\sigma}_{(s,1)}$ is within a known interval of length $P$. Then $\boldsymbol{\sigma}_{(s,1)}$ can be recovered and the exact erroneous coordinate of $\boldsymbol{\sigma}_{(s,1)}$ can be located.
\end{lemma}

\begin{IEEEproof}
Suppose $\boldsymbol\sigma\in \mathrm{PSVT}_{b_1,b_2,c_1,c_2}(n,s,P)$ suffers a burst stable deletion of length $s$ and the erroneous coordinate of $\boldsymbol{\sigma}_{(s,1)}$ is within a known interval of length $P$. Since $\boldsymbol{\sigma} \in \mathcal{C}_{c_1,c_2}^{\mathrm{retrieve}}(n,s,P)$, by Theorem \ref{thm:symbol} we can retrieve the missing symbol of $\boldsymbol{\sigma}_{(s,1)}$. By an $n$-ary SVT decoder, the erroneous coordinate of $\alpha(\boldsymbol{\sigma}_{(s,1)})$ lies in a consecutive run of 0's or 1's, which means that the erroneous coordinate of $\boldsymbol{\sigma}_{(s,1)}$ lies in a monotone decreasing sequence or a monotone increasing sequence. In such a sequence, the coordinate of the known missing symbol is certainly unique. Therefore, the erroneous coordinate of $\boldsymbol{\sigma}_{(s,1)}$ can be exactly located.
\end{IEEEproof}

We make use of the PSVT codes above for $2\leq s' \leq s$ to finally construct our $^\leq s$-BSD permutation codes as follows.

\begin{construction}
Assume that $\frac{n}{s}$ is even and $n$ is a multiple of $2 \prod_{s'=1}^s s' P_{s'}$ where $P_{s'}= \lceil (\delta+s'-1)/s' \rceil$, $1\leq s' \leq s$.
Let $\boldsymbol{a}=(a_1,a_2)$ where $a_1 \in \mathbb{Z}_4, a_2 \in \mathbb{Z}_{2n}$.
Let $\boldsymbol{b}=(b_1,b_2,\dots,b_{2s})$ where $b_{2s'-1} \in \mathbb{Z}_{P_{s'}}$, $b_{2s'} \in \mathbb{Z}_2$ for $1 \leq s' \leq s$. Let $\boldsymbol{c}=(c_3,c_4,\dots,c_{2s})$ where $c_{2s'-1}, c_{2s'} \in \mathbb{Z}_{2s' P_{s'}}$ for $2 \leq s' \leq s$.
Let $\boldsymbol{d}=(d_1,d_2)$ where $d_1, d_2 \in \mathbb{Z}_{(4s)!}$.
Define the code $\mathcal{C}_s^3 \triangleq \mathcal{C}_s^3 (n;\boldsymbol{a};\boldsymbol{b},\boldsymbol{c};\boldsymbol{d})$ as
\begin{align*}
    \Big\{ \boldsymbol{\sigma} \in S_n:
    & ~ \mathbbm{1}(\boldsymbol{\sigma}) \in \mathcal{C}_{a_1,a_2}^{\mathrm{locate}}(n,s), \\
    & \alpha(\boldsymbol{\sigma}) \in \mathrm{SVT}_{b_1,b_2}(n,P_1),\\
    & \boldsymbol{\sigma} \in \mathcal{C}_{d_1,d_2}^{\mathrm{recover}}(n,2s), \\
    & \boldsymbol{\sigma} \in \bigcap_{s'=2}^s \mathrm{PSVT}_{b_{2s'-1}, b_{2s'}, c_{2s'-1}, c_{2s'}}(n,s',P_{s'}) \Big\}.
\end{align*}
\end{construction}

\begin{theorem}\label{thm:varperm}
The permutation code $\mathcal{C}_s^3$ is an $^{\leq} s$-BSD permutation code over $S_n$.
Moreover, by choosing proper parameters there is an $^{\leq} s$-BSD permutation code with redundancy at most
$\log  n + (3s-2) \log  \log  n +O(1)$.
\end{theorem}

\begin{IEEEproof}
Suppose $\boldsymbol{\sigma} = \left( \sigma_1,\sigma_2,\ldots,\sigma_n \right) \in \mathcal{C}_s^3 (n;\boldsymbol{a};\boldsymbol{b},\boldsymbol{c};\boldsymbol{d})$ suffers a burst stable deletion of length $s'$ where $1 \leq s' \leq s$. Since its parity indicator vector $\mathbbm{1}(\boldsymbol{\sigma}) \in \mathcal{C}_{a_1,a_2}^{\mathrm{locate}}(n,s)$, by Lemma \ref{lem:dense} all erroneous coordinates can be located within an interval of length at most $\delta+s'-1$.
Now we view $\boldsymbol{\sigma}$ in an $s'\times \frac{n}{s'}$ array, and then the erroneous coordinate in the first row of such an array is located in an interval of length at most $P_{s'}$.

For $s'=1$, the unique missing symbol of a permutation can be trivially retrieved. Since $\alpha(\boldsymbol{\sigma}) \in \mathrm{SVT}_{b_1,b_2}(n,P_1)$, we can recover $\boldsymbol{\sigma}$ by an $n$-ary SVT decoder.

For $2 \leq s' \leq s$, $\boldsymbol{\sigma} \in \mathrm{PSVT}_{b_{2s'-1}, b_{2s'}, c_{2s'-1}, c_{2s'}}(n,s',P_{s'})$. By viewing $\boldsymbol{\sigma}$ in an $s'\times \frac{n}{s'}$ array, in the first row $\boldsymbol{\sigma}_{(s',1)}$ the erroneous coordinate is within a known interval of length $P_{s'}$. By Lemma \ref{lem:PSVT} and Theorem \ref{thm:symbol}, $\boldsymbol{\sigma}_{(s',1)}$ can be recovered and the erroneous coordinate of $\boldsymbol{\sigma}_{(s',1)}$ is exactly located, and thus all erroneous coordinates of $\boldsymbol{\sigma}$ are located in an interval of length $2s'$. Then we view $\boldsymbol{\sigma}$ in an $2s\times \frac{n}{2s}$ array and all erroneous coordinates can be located in two adjacent columns in this array. Finally, since $\boldsymbol{\sigma} \in \mathcal{C}_{d_1,d_2}^{\mathrm{recover}}(n,2s)$, the correction of the two columns can be done in a similar way as Lemma \ref{lem:recover}.

By the pigeon-hole principal, there exist parameters $a_1 \in \mathbb{Z}_{4}$, $a_2 \in \mathbb{Z}_{2n}$, $b_{2s'-1} \in \mathbb{Z}_{P_{s'}}$, $b_{2s'} \in \mathbb{Z}_2$ for $1 \leq s' \leq s$, $c_{2s'-1}, c_{2s'} \in \mathbb{Z}_{2s'P_{s'}}$ for $2 \leq s' \leq s$, and $d_1,d_2 \in \mathbb{Z}_{(4s)!}$, such that the redundancy of $\mathcal{C}_s^3$ is at most
\begin{align*}
&
\begin{aligned}
\log  2 + \log  4 + \log  2n + \sum_{s'=1}^s (\log  P_{s'} + \log   2) \\
+ \sum_{s'=2}^s 2 \log  2s'P_{s'}+  2 \log  (4s)!
\end{aligned}\\
= &\log  n+ (3s-2)\log \log  n+O(1),
\end{align*}
where the first term is the 1 bit redundancy due to the size of $(\boldsymbol{p},\delta)$-dense permutations (Lemma \ref{lem:densesize}).
\end{IEEEproof}

\subsection{$^\leq s$-BSD multi-permutation codes}\label{subsec:varmulti}

Due to the existence of repeated entries in a multi-permutation, we need a little modification as we do in the previous section. Given $s$ and $P$, we define the MPSVT codes as follows (which means the SVT version for multi-permutations).

\begin{construction}\label{con:MPSVT}
Let $\boldsymbol{r} = \left( r_1,r_2,\ldots,r_w \right)$ be a multiplicity vector. Let $b_1 \in \mathbb{Z}_{P}$, $b_2 \in \mathbb{Z}_2$, and $c_1,c_2 \in \mathbb{Z}_{2Ps}$, define the code $\mathrm{MPSVT}_{b_1,b_2,c_1,c_2}(n,s,P)$ as
    \begin{align*}
        \big\{ \boldsymbol{\sigma} \in S_n^{\boldsymbol{r}}:
        &~ \boldsymbol{\sigma} \in \mathcal{C}_{c_1,c_2}^{\mathrm{retrieve}}(n,s,P), \\
        & \alpha(\boldsymbol{\sigma}_{(s,1)}) \in \mathrm{SVT}_{b_1,b_2}(n,P) \big\}.
    \end{align*}
\end{construction}

\begin{lemma}\label{lem:MPSVT}
Suppose an arbitrary multi-permutation $\boldsymbol\sigma\in \mathrm{MPSVT}_{b_1,b_2,c_1,c_2}(n,s,P)$ suffers a burst stable deletion of length $s$ and the erroneous coordinate of $\boldsymbol{\sigma}_{(s,1)}$ is within a known interval of length $P$. Then $\boldsymbol{\sigma}_{(s,1)}$ can be recovered and the exact erroneous coordinate of $\boldsymbol{\sigma}_{(s,1)}$ can be located within an interval of length $r$ in the worst case, where $r= \max_{1 \leq i \leq w} r_i$.
\end{lemma}

Lemma \ref{lem:MPSVT} can be proved analogously as Lemma \ref{lem:PSVT}, with the only difference in the final step due to the existence of repeated entries.
We make use of the MPSVT codes above for $2\leq s' \leq s$ to finally construct our $^\leq s$-BSD multi-permutation codes as follows, assuming the multiplicity vector $\boldsymbol{r}$ is regular.

\begin{construction}\label{con:varmultiperm}
Assume $n=rw=s(r+1)t$ where $t$ is even.
Further we assume that $n$ is a multiple of $2 \prod_{s'=1}^s s' P_{s'}$ where $P_{s'}= \lceil (\delta+s'-1)/s' \rceil$, $1\leq s' \leq s$.
Let $\boldsymbol{a}=(a_1,a_2)$ where $a_1 \in \mathbb{Z}_4, a_2 \in \mathbb{Z}_{2n}$.
Let $\boldsymbol{b}=(b_1,b_2,\dots,b_{2s})$ where $b_{2s'-1} \in \mathbb{Z}_{P_{s'}}$ and $b_{2s'} \in \mathbb{Z}_2$ for $1 \leq s' \leq s$. Let $\boldsymbol{c}=(c_3,c_4,\dots,c_{2s})$ where $c_{2s'-1}, c_{2s'} \in \mathbb{Z}_{2s'P_{s'}}$ for $2 \leq s' \leq s$.
Let $\boldsymbol{d}=(d_1,d_2)$ where $d_1, d_2 \in \mathbb{Z}_{(2s(r+1))!}$.
Define the code $\mathcal{C}_s^4 \triangleq \mathcal{C}_s^4 (n;\boldsymbol{a};\boldsymbol{b},\boldsymbol{c};\boldsymbol{d})$ as
\begin{align*}
    \Big\{ \boldsymbol{\sigma} \in S_n^{\boldsymbol{r}}:
    & ~ \mathbbm{1}(\boldsymbol{\sigma}) \in \mathcal{C}_{a_1,a_2}^{\mathrm{locate}}(n,s), \\
    & \alpha(\boldsymbol{\sigma}) \in \mathrm{SVT}_{b_1,b_2}(n,P_1),\\
    & \boldsymbol{\sigma} \in \mathcal{C}_{d_1,d_2}^{\mathrm{recover}}(n,s(r+1)), \\
    & \boldsymbol{\sigma} \in \bigcap_{s'=2}^s \mathrm{MPSVT}_{b_{2s'-1}, b_{2s'}, c_{2s'-1}, c_{2s'}}(n,s',P_{s'}) \Big\}.
\end{align*}
\end{construction}

\begin{theorem}\label{thm:varmultiperm}
 The code $\mathcal{C}_s^4$ we constructed in Construction \ref{con:varmultiperm} is an $^{\leq }s$-BSD multi-permutation code over $S_n^{\boldsymbol{r}}$ ($\boldsymbol{r}$ is a regular multiplicity vector). Moreover, by choosing proper parameters there is an $^{\leq }s$-BSD multi-permutation code over $S_n^{\boldsymbol{r}}$ with redundancy at most $\log  n + (3s-2) \log  \log  n +O(1)$.
\end{theorem}

\begin{IEEEproof}
The proof works in the same way as the proof of Theorem \ref{thm:varperm} until we recover $\boldsymbol{\sigma}_{(s',1)}$.
However, due to repeated entries in multi-permutations, here we can only locate the erroneous coordinate of $\boldsymbol{\sigma}_{(s',1)}$ in an interval of length $r$, in the worst case. Thus, all erroneous coordinates of $\boldsymbol{\sigma}$ are located in an interval of length $s'(r+1)$. By viewing $\boldsymbol{\sigma}$ in an array of size $s(r+1)\times \frac{n}{s(r+1)}$, all the erroneous coordinates are within two adjacent columns of this array. Finally, since $\boldsymbol{\sigma} \in \mathcal{C}_{d_1,d_2}^{\mathrm{recover}}(n,s(r+1))$, the correction of the two columns can be done in a similar way as Lemma \ref{lem:recover}.

The code redundancy can be calculated by the pigeon-hole principle immediately.
\end{IEEEproof}

To construct codes for irregular multi-permutations, simply choose $r_{\mathrm{max}} = \max_{i \in [w]} r_i$ and apply the above construction.

\begin{corollary}
There exists an $^{\leq} s$-BSD multi-permutation code in $S_n^{\boldsymbol{r}}$ ($\boldsymbol{r}$ could be irregular) with redundancy at most $\log  n + (3s-2) \log  \log  n +O(1)$.
\end{corollary}

\section{A linear-time encoder for single stable deletion correcting permutation codes}\label{Sec:Enc}

Previous sections consist of the stable deletion correcting permutation or multi-permutation codes, together with their decoding algorithms.
In this section, we consider the problem of designing an efficient encoder that encodes a user message, denoted by an arbitrary permutation $\boldsymbol{\pi} \in S_{k}$,
into a permutation lying in a stable deletion correcting permutation code $\mathcal{C}\subseteq S_n$.

It should be noted that designing efficient encoders in deletion error models is quite a challenging task.
For the binary case, the encoder proposed by Abdel-Ghaffar and Ferreira \cite{Abdel-Ghaffar-98-IT-Algorithm} with almost optimal redundancy for the binary VT code
did not appear until 1998. For the $q$-ary case, so far there is no efficient encoder for the $q$-ary VT code $\mathrm{VT}_{a,b}(n,q)$ matching the optimal redundancy, and the existential best known encoder proposed by Abroshan et al. \cite{Abroshan-18-ISIT-Algorithm} has redundancy $(\log q + 1)\lceil \log n \rceil + 2(\log q - 1)$. A very recent result \cite{Nguyen-22-arXiv-Algorithm} introduces a variant of $q$-ary VT code with more efficient encoder.

To the best of our knowledge, so far there is no research on efficient encoders for deletion correcting permutation codes, even for only a single deletion.
Recall that the code $\mathcal{C}_a(n) = \{\boldsymbol{\sigma} \in S_n : \alpha(\boldsymbol{\sigma}) \in \mathrm{VT}_a(n-1) \}$, where $0 \leq a < n$, is a single stable deletion correcting permutation code given by Levenshtein \cite{Levenshtein-92-DM-ID}. In this section, we give a linear-time encoder which encodes an arbitrary permutation $\boldsymbol{\pi} \in S_{n-1}$ into a permutation in $\mathcal{C}_a(n)$ and thus the encoder has the optimal redundancy $\log n$. The key idea is to show that we can simply find an appropriate coordinate to insert the symbol $n$ and guarantee that the resultant permutation indeed belongs to $\mathcal{C}_a(n)$. The encoder and the proof of its correctness are as follows.

\begin{algorithm}
    \caption{Single stable deletion correcting permutation codes encoder}\label{encoder}
    \KwIn{$\boldsymbol{\pi} \in S_{n-1}$ and $0 \leq a < n$}
    \KwOut{$\boldsymbol{\sigma}= (\pi_1, \cdots, \pi_{j}, n, \pi_{j+1}, \cdots, \pi_{n-1}) \in \mathcal{C}_a(n)$}

    \textbf{Initialization:} Let $\Delta= (a- \sum_{i=1}^{n-2}i \alpha(\pi_i) ) \pmod{n}$ and $\omega= \sum_{i=1}^{n-2} \alpha(\pi_i)$\\
        \If{$0 \leq \Delta \leq \omega-1$}{
            Let $j$ be the coordinate of the $(\omega- \Delta)$-th $1$ in $\alpha(\boldsymbol{\pi})$}

        \If{$\Delta= \omega$}
            {
                Let $j=0$
            }

        \If{$\omega<\Delta<n-1$}
            {
                Let $j$ be the coordinate of $(\Delta-\omega)$-th $0$ in $\alpha(\pi)$
            }

        \If{$\Delta=n-1$}
            {
                Let $j=n-1$
            }
\end{algorithm}

\begin{theorem}
  Algorithm \ref{encoder} is correct and runs in $O(n)$ time.
\end{theorem}

\begin{IEEEproof}
    For any $\boldsymbol{\pi} \in S_{n-1}$, let $\Delta= (a- \sum_{i=1}^{n-2}i \alpha(\pi_i) ) \pmod{n}$ and $\omega= \sum_{i=1}^{n-2} \alpha(\pi_i)$.  Then we have $0 \leq \Delta \leq n-1$ and $0 \leq \omega \leq n-2$.

    If $0 \leq \Delta \leq \omega-1$, let $j$ be the coordinate of the $(\omega- \Delta)$-th $1$ in $\alpha(\boldsymbol{\pi})$. Then, $\alpha(\pi_j)=1$, i.e., $\pi_j \leq \pi_{j+1}$, and we insert $n$ between $\pi_{j}$ and $\pi_{j+1}$, i.e., $\boldsymbol{\sigma}= (\pi_1, \cdots, \pi_{j}, n, \pi_{j+1}, \cdots, \pi_{n-1})$. We obtain
    \begin{align*}
    \alpha(\boldsymbol{\sigma})
    &= (\alpha(\pi_1), \cdots, \alpha(\pi_{j-1}), 1, 0, \alpha(\pi_{j+1}), \cdots, \alpha(\pi_{n-2})) \\
    &= (\alpha(\pi_1), \cdots, \alpha(\pi_{j}), 0, \alpha(\pi_{j+1}), \cdots, \alpha(\pi_{n-2})),
    \end{align*}
    that is, $\alpha(\boldsymbol{\sigma})$ is obtained by inserting a symbol $0$ at the $j$-th coordinate in $\alpha(\boldsymbol{\pi})$. Now, we have
          \begin{align*}
            \sum_{i=1}^{n-1}i \alpha(\sigma_i)
            &= \sum_{i=1}^{j}i \alpha(\pi_i) + \sum_{i=j+2}^{n-1}i \alpha(\pi_{i-1}) \\
            &= \sum_{i=1}^{n-2}i \alpha(\pi_i) + \sum_{i=j+1}^{n-2}\alpha(\pi_{i}) \\
            &= (a - \Delta)  + \omega- (\omega-\Delta) \\
            &= a \pmod{n},
          \end{align*}
          and thus indeed we have $\boldsymbol{\sigma} \in \mathcal{C}_a(n)$.

    If $\Delta= \omega$, we insert $n$ at the very beginning of $\boldsymbol{\pi}$, that is, $\boldsymbol{\sigma}= (n, \pi_1, \cdots, \pi_{n-1})$, and we obtain $\alpha(\boldsymbol{\sigma})= (0, \alpha(\pi_1), \cdots, \alpha(\pi_{n-2}))$, that is, $\alpha(\boldsymbol{\sigma})$ is obtained by inserting a symbol $0$ at the very beginning of $\alpha(\boldsymbol{\pi})$. Now, we have
          \begin{align*}
            \sum_{i=1}^{n-1}i \alpha(\sigma_i)
            &= \sum_{i=2}^{n-1} i \alpha(\pi_{i-1})\\
            &= \sum_{i=1}^{n-2}i \alpha(\pi_i) + \sum_{i=1}^{n-2}\alpha(\pi_{i}) \\
            &= (a - \Delta)  + \Delta \\
            &= a \pmod{n},
          \end{align*}
          and thus indeed we have $\boldsymbol{\sigma} \in \mathcal{C}_a(n)$.

      If $\omega<\Delta<n-1$, let $j$ be the coordinate of $(\Delta-\omega)$-th $0$ in $\alpha(\boldsymbol{\pi})$. Note that $\alpha(\boldsymbol{\sigma})$ has $n-2-\omega$ zeros and thus such a coordinate must exist. Then, $\alpha(\pi_j)=0$, i.e., $\pi_j > \pi_{j+1}$, and we insert $n$ between $\pi_{j}$ and $\pi_{j+1}$, i.e., $\boldsymbol{\sigma}= (\pi_1, \cdots, \pi_{j}, n, \pi_{j+1}, \cdots, \pi_{n-1})$. We obtain
      \begin{align*}
      &~~~~ \alpha(\boldsymbol{\sigma})\\
      &= (\alpha(\pi_1), \cdots, \alpha(\pi_{j-1}), 1, 0, \alpha(\pi_{j+1}), \cdots, \alpha(\pi_{n-2})) \\
      &= (\alpha(\pi_1), \cdots, \alpha(\pi_{j-1}), 1, \alpha(\pi_j), \alpha(\pi_{j+1}), \cdots, \alpha(\pi_{n-2})),
      \end{align*}
      that is, $\alpha(\boldsymbol{\sigma})$ is obtained by inserting a symbol $1$ at the $(j-1)$-th coordinate in $\alpha(\boldsymbol{\pi})$. Now, we have
          \begin{align*}
            \sum_{i=1}^{n-1}i \alpha(\sigma_i)
            &= \sum_{i=1}^{j-1}i \alpha(\pi_i) + j + \sum_{i=j+1}^{n-1}i \alpha(\pi_{i-1}) \\
            &= \sum_{i=1}^{n-2}i \alpha(\pi_i) + j + \sum_{i=j}^{n-2}\alpha(\pi_{i}) \\
            &
            \begin{aligned}
             =(a-\Delta)
             &+ j+ (n-2-j+1) \\
             &-(n-2-\omega- (\Delta- \omega- 1))
            \end{aligned} \\
            &= a \pmod{n},
          \end{align*}
          and thus indeed we have  $\boldsymbol{\sigma} \in \mathcal{C}_a(n)$.

    Finally, if $\Delta= n-1$, we insert $n$ at the very end of $\boldsymbol{\pi}$, that is, $\boldsymbol{\sigma}= (\pi_1, \cdots, \pi_{n-1},n)$. We obtain $\alpha(\boldsymbol{\sigma})= ( \alpha(\pi_1), \cdots, \alpha(\pi_{n-2}),1)$, that is, $\alpha(\boldsymbol{\sigma})$ is obtained by inserting a symbol $1$ at the very end of $\alpha(\boldsymbol{\pi})$. Now, we have
          \begin{align*}
            \sum_{i=1}^{n-1}i \alpha(\sigma_i)
            &= \sum_{i=1}^{n-2} i \alpha(\pi_{i}) + n-1 \\
            &= (a - \Delta)  + n-1 \\
            &= a \pmod{n},
          \end{align*}
          and thus indeed we have $\boldsymbol{\sigma} \in \mathcal{C}_a(n)$.

    To sum up, for each of the four cases, we encode $\boldsymbol{\pi} \in S_{n-1}$ as $\boldsymbol{\sigma} \in \mathcal{C}_a(n)$. Furthermore, it is straightforward to check that we can calculate $\Delta$ and $\omega$ and find $j$ in $O(n)$ time.
\end{IEEEproof}

\begin{example}
  Let $n=10$ and $a=0$.
  \begin{itemize}
    \item If $\boldsymbol{\pi}= (2,1,4,3,6,5,8,7,9)$, then we have $\alpha(\boldsymbol{\pi})= (0,1,0,1,0,1,0,1)$. Now, let $\Delta= (0-\sum_{i=1}^8 i\alpha(\pi_i)) \pmod{10}= 0$ and $\omega= \sum_{i=1}^8 \alpha(\pi_i)= 4$. Then, $0 \leq \Delta \leq \omega-1$, the coordinate of $4$-th $1$ ($\omega-\Delta=4$) in $\alpha(\boldsymbol{\pi})$ is $8$, and we let $\boldsymbol{\sigma}= (2,1,4,3,6,5,8,7,10,9)$. Thus, we obtain $\alpha(\boldsymbol{\sigma})= (0,1,0,1,0,1,0,1,0)$ and $\sum_{i=1}^9 i \alpha(\sigma_i)= 0 \pmod{10}$.
    \item If $\boldsymbol{\pi}= (1,2,4,3,9,8,7,6,5)$, then we have $\alpha(\boldsymbol{\pi})= (1,1,0,1,0,0,0,0)$. Now, let $\Delta= (0-\sum_{i=1}^8 i\alpha(\pi_i)) \pmod{10}= 3$ and $\omega= \sum_{i=1}^8 \alpha(\pi_i)= 3$. Then $\Delta= \omega$ and we let $\boldsymbol{\sigma}= (10,1,2,4,3,9,8,7,6,5)$. Thus, we obtain that $\alpha(\boldsymbol{\sigma})= (0,1,1,0,1,0,0,0,0)$ and $\sum_{i=1}^9 i \alpha(\sigma_i)= 0 \pmod{10}$.
    \item If $\boldsymbol{\pi}= (3,1,2,9,8,7,6,5,4)$, then we have $\alpha(\boldsymbol{\pi})= (0,1,1,0,0,0,0,0)$. Now, let $\Delta= (0-\sum_{i=1}^8 i\alpha(\pi_i)) \pmod{10}= 5$ and $\omega= \sum_{i=1}^8 \alpha(\pi_i)= 2$. Then $\omega<\Delta<n-1$, the coordinate of $3$-th $0$ ($\Delta-\omega=3$) in $\alpha(\boldsymbol{\pi})$ is $5$, and we let $\boldsymbol{\sigma}= (3,1,2,9,8,10,7,6,5,4)$. Now, we have $\alpha(\boldsymbol{\sigma})= (0,1,1,0,1,0,0,0,0)$ and $\sum_{i=1}^9 i \alpha(\sigma_i)= 0 \pmod{10}$.
    \item If $\boldsymbol{\pi}= (1,9,8,7,6,5,4,3,2)$, then we have $\alpha(\boldsymbol{\pi})= (1,0,0,0,0,0,0,0)$. Now, let $\Delta= (0-\sum_{i=1}^8 i\alpha(\pi_i)) \pmod{10}= 9= n-1$. Then we let the permutation $\boldsymbol{\sigma}= (1,9,8,7,6,5,4,3,2,10)$. Thus, we obtain that $\alpha(\boldsymbol{\sigma})= (1,0,0,0,0,0,0,0,1)$ and $\sum_{i=1}^9 i \alpha(\sigma_i)= 0 \pmod{10}$.

  \end{itemize}
\end{example}

Moreover, our encoding algorithm clearly suggests a rather simple decoding algorithm, presented in Algorithm \ref{decoder}.

\begin{algorithm}[!htb]
    \caption{Single stable deletion correcting permutation code decoder}\label{decoder}
    \KwIn{ $\boldsymbol{\sigma}' \in \mathcal{B}_1(\boldsymbol{\sigma})$ where $\boldsymbol{\sigma}\in \mathcal{C}_a(n)$}
    \KwOut{$\boldsymbol{\pi} \in S_{n-1}$}

    \textbf{Step 1:} Recover $\boldsymbol{\sigma}$ from $\boldsymbol{\sigma}'$\\

    \textbf{Step 2:} Find the symbol $n$ in $\boldsymbol{\sigma}$ and remove it\\
\end{algorithm}

%\begin{remark}
%  For multi-permutations, let $\boldsymbol{r}= (r_1,r_2,\ldots,r_w)$ be a multiplicity vector where $\sum_{i=1}^{w}r_i= n-1$,
%  Algorithm \ref{encoder} works when we encode a multi-permutation $\boldsymbol{\pi} \in S_{n-1}^{\boldsymbol{r}}$ into $\boldsymbol{\sigma} \in \mathcal{C}_a(n) \subseteq S_{n}^{\boldsymbol{r}'}$ where $\boldsymbol{r}'= (r_1,r_2,\ldots,r_w,r_{w+1})$ and $r_{w+1}=1$.
%\end{remark}

\section{Conclusion}\label{Sec:concl}

Motivated by applications in flash memories, in this paper we consider permutation codes and multi-permutation codes against burst stable deletions.
Our main improvement relies on a different approach to retrieve the missing symbol on the first row of the array representation of a permutation or a multi-permutation, with less redundancy. The gap between the redundancy of our codes and the theoretic bound is only of order $\log \log n$. The more difficult model of burst unstable  deletions for permutations and multi-permutations are considered for future research. Another interesting problem is to investigate the efficient encoder for burst stable deletion correcting permutation and multi-permutation codes.

\end{document}